\documentclass[a4paper,12pt]{article}

\pdfoutput=1 

\usepackage[english]{babel}
\usepackage{graphicx,subfig}
\usepackage{a4wide}
\usepackage{amsfonts,amsmath,amssymb,bm}
\usepackage{slashed}
\usepackage[final,bookmarks]{hyperref}
\usepackage{color,comment}

\newcommand{\eq}{ \; = \; }

\newcommand{\dirac}{ \delta_{D}^{(3)} }
\newcommand{\hubble}{ \mathcal{H} }
\newcommand{\kNL}{ k_\mathrm{NL} }
\newcommand{\EV}[2]{ {\big\langle} #1 \, #2 {\big\rangle} }

\newcommand{\intd}[1]{\int d^{3}#1 \, }
\newcommand{\vx}{{\bf x}}
\newcommand{\vk}{{\bf k}}
\newcommand{\vy}{{\bf y}}
\newcommand{\vw}{{\bm \omega}}
\newcommand{\vq}{{\bf q}}
\newcommand{\vv}{{\bf v}}
\newcommand{\vxB}{{\bf x}_{B}}
\newcommand{\vp}{{\bm \pi}}
\newcommand{\vn}{{\bm \nu}}
\newcommand{\vm}{{\bm \mu}}

\newcommand{\ex}[1]{\langle #1 \rangle}

\newcommand{\va}{\bm{ \alpha}_{\omega}}
\newcommand{\vb}{\bm \beta_{\omega}}
\newcommand{\DB}{\Delta_{B}}
\newcommand{\vxu}{{\bf x}^{(1)}}
\newcommand{\vvu}{{\bf v}^{(1)}}
\newcommand{\vxt}{{\bf x}^{(2)}}
\newcommand{\vvt}{{\bf v}^{(2)}}
\newcommand{\WL}{W_{\Lambda}}
\newcommand{\vu}{{\bf u}}

\newcommand{\be}{\begin{eqnarray} }
\newcommand{\ee}{\end{eqnarray} }
\newcommand{\bs}{\begin{split} }
\newcommand{\es}{\end{split} }
\newcommand{\inv}{\mathcal{I}}

\newcommand{\cH}{ \mathcal{H} }
\newcommand{\non}{\nonumber \\}
\newcommand{\mb}{ \beta^{\omega\omega}}
\newcommand{\mg}{ \gamma^{\omega}}
\newcommand{\vg}{ \gamma^{\omega\omega}}

\newcommand{\tJ}{\tilde J}
\newcommand{\dss}{d^{2}}
\newcommand{\dst}{\tilde d^{2}}

\newcommand{\intq}{\int_{\bf q}}

\newcommand{\intqp}{\int_{\bf q'}}

\newcommand{\vvol}{\vec v_{\rm vol}}
\newcommand{\vdis}{\vec v_{\rm disc}}
\newcommand{\vmas}{\vec  v_{\rm mass}}
\newcommand{\smo}[1]{ \left[ #1 \right]_{\Lambda}}

\renewcommand\vec[1]{{\bf #1}}

\allowdisplaybreaks[1]

\begin{document}


\thispagestyle{empty}  
\setcounter{page}{0}


\begin{center}
{\huge On the Velocity in the\\[0.5ex] Effective Field Theory of Large Scale \\[0.5ex] Structures}\\[1.6cm]

\textbf{Lorenzo Mercolli$^a$} and \textbf{Enrico Pajer$^b$} \\[3ex]

$^a$  \emph{Department of Astrophysical Sciences, Princeton University, \\ Princeton, NJ 08544, USA }\\[3ex]
$^b$  \emph{Department of Physics, Princeton University, \\ Princeton, NJ 08544, USA }\\[9ex]

\textbf{Abstract}
\end{center}
\begin{quote}
\noindent We compute the renormalized two-point functions of density, divergence and vorticity of the velocity in the Effective Field Theory of Large Scale Structures. 
Because of momentum and mass conservation, the corrections from short scales to the large-scale power spectra of density, divergence and vorticity must start at order $k^{4}$. For the vorticity this constitutes one of the two leading terms. Exact (approximated) self-similarity of an Einstein-de Sitter ($\Lambda$CDM) background fixes the time dependence so that the vorticity power spectrum at leading order is determined by the symmetries of the problem and the power spectrum around the non-linear scale. We show that to cancel all divergences in the velocity correlators one needs new counterterms. These fix the definition of velocity and do not represent new properties of the system. For an Einstein-de Sitter universe, we show that all three renormalized cross- and auto-correlation functions have the same structure but different numerical coefficients, which we compute.  We elucidate the differences between using momentum and velocity.


\end{quote}

\newpage

\tableofcontents

\section{Introduction}

In the era of large cosmological surveys and numerical simulations the need for simple and intuitive analytical guidance is stronger than ever. This is especially true in the study of the large scale structures (LSS) of the universe at late times where the validity of simple linear theory is much more limited than in the early universe. Of the many issues encountered in trying to understand and observationally test LSS, such as for example biasing, baryonic physics, redshift space distortion, relativistic effects etc., one seems to be particularly suited for analytical investigation: the evolution of the distribution of dark matter. We have certainly many tools to deal with many-body problems, but we have to be careful in dealing with the long-range attractive gravitational interaction. Because of gravity, scales much larger than the Hubble horizon have been frozen since the very early universe. At these scales (today of the order of 10 Gpc) the universe is homogeneous to a part in 100,000 and so linear theory is very accurate. But gravity also causes the growth of inhomogeneities with time, pushing scales shorter than a few Mpc into a completely non-linear regime. Considering that we are powerless in describing analytically realistic non-linear gravitational collapse and that short scales couple gravitationally to large scales, one might abandon oneself to pessimism. Fortunately, Effective Field Theories (EFT) are designed to tackle precisely these problems, namely when things we care about interact with others we do not know or care about (see e.g.~Ref.~\cite{Weinberg1996a} for a general introduction). Applications of EFT to the well developed theory of LSS, nicely reviewed in the standard reference \cite{Bernardeau2002}, have recently attracted growing interest \cite{Baumann2012,Carrasco2012,Hertzberg2012,Pajer2013,Carrasco2013}. The key point of this approach, hereafter EFTofLSS, is that short scales, which are traditionally neglected, affect large scales in a specific and predictable manner. The only way to set up a consistent perturbation theory is to systematically account for these short-scale effects. In practice, this amounts to including in the EFTofLSS all operators compatible with the symmetries of the problem as we will review in Sec.~\ref{s:1}. When neglecting these terms, one loses the ability to renormalize the theory. 

At this point, the sceptical reader will object: ``But LSS are well described by a classical, deterministic theory for which there is no need for renormalization''. Things are in fact different for the following reason. This particular realization of the universe in which we live can indeed be described at cosmological scales by simply evolving some specific initial conditions with some classical deterministic set of equations. But as of now, we have no way to know what these specific initial conditions were. What we do is to average over some ensemble of initial conditions and invoke the ergodic theorem to compare those ensemble averages to spatial averages of what we observe around us. It is because of the ensemble average that our description of LSS is not a deterministic one. Instead it is analogous to a statistical (or quantum mechanical) description and hence the field theory we are using comes with all the interesting phenomena observed in those fields. In particular, renormalization is not something we do to get nicer fit from theory, instead it is a crucial step in the \textit{definition} of a consistent theory, as for any statistical or quantum field theory.

The sceptical reader might object again: ``In $\Lambda$CDM the loops arising in perturbation theory are not divergent so there is nothing to renormalize''. Actually, renormalization is not really about cancelling infinities, rather it is about handling those situations in which the momentum of some internal integral, i.e.~a loop\footnote{Just to be clear, there is nothing quantum about these loops. Remember that loops, i.e.~integrals over internal momenta, arise because we are solving perturbatively a non-linear theory. A local quadratic interaction such as $\delta(x)\theta(x)$ becomes a convolution in momentum space, hence the momentum integral.}, takes values that are much larger than those for which we trust the theory under consideration. We will call these contributions ``UV-corrections'', since it is  in the limit of high energies and large momenta that the EFTofLSS ceases to be valid. The concrete value of these UV-corrections, whether finite or not, are meaningless and cannot be trusted. Renormalization is the process of defining a limit in which the sum of UV-correction and appropriate counterterms gives a finite physical value\footnote{The choice of renormalization scheme and the related renormalization group equations are exactly about the arbitrariness in distinguishing between UV-corrections and counterterms.}. Because of this, whether or not $\Lambda$CDM (or any other solution of the theory) has finite or infinite UV-corrections, the field theory for LSS have to be renormalized in order to give physical results.  

In this paper we make progress in the study of the EFTofLSS in several directions. For the reader that is not interested in the details of the calculation or that cannot wait any longer to hear our results, we present here a summary of our main findings.

\paragraph{Velocity and mass and momentum conservation} Besides the momentum $\vp$, the other natural variable to use in the EFTofLSS is the mass-weighted velocity, i.e.~$\vv_{l}\sim\vp_{l}/\rho_{l}+\rm{counterterms}$  (this relation must include appropriate counterterms for the renormalization of composite operators as we discuss in subsection \ref{s:sims}), where $\rho$ is the density and the label $l$ indicates that a quantity contains only large-scale modes. 
This is to be contrasted with the volume-weighted definition $ \tilde \vv_{l} \equiv\left(\vp/\rho\right)_{l}$. For the mass-weighted velocity it is possible to construct an argument similar to the one reviewed in chapter 28 of Ref.~\cite{Peebles1980} and originally due to Y.~B.~Zel'dovich \cite{Zeldovich1965}. One finds that \textit{the effect of short scales on the power spectrum $P$ of the divergence $\theta\equiv\nabla\cdot \vv_{l}$ and vorticity $\bm{\omega}_l\equiv\nabla \times \vv_l$ of the velocity must go to zero as $P\sim k^{4}$ in the limit $k\rightarrow 0$}. Besides explaining why there are exact cancellations of the $k^{2}$ terms in $P_{22}$ of both $\delta$ and $\theta$ this argument has interesting implications for the vorticity\footnote{Notice that the consequences of mass and momentum conservation for $\delta$ and $\theta$ (e.g.~the fact that the SPT kernels start with $k^{2}$ in the $k\rightarrow 0$ limit) had long been known \cite{Goroff1986} (see also \cite{Crocce2008} for a confirmation using simulations). On the other hand, to the best of our knowledge, the implications for the vorticity have not yet been discussed in the literature.} (see below).

\paragraph{Vorticity} Although the vorticity decays as $\bm{\omega }\sim a^{-1}$ in linear theory, it gets sourced by the EFT corrections, most importantly by a stochastic noise term and from higher order terms in the stress tensor that arise from integrating out short scales. In an Einstein de Sitter (EdS) universe with power-law initial conditions $P_{\delta\delta}\sim k^{n}$, these two terms give the two leading contributions to the vorticity power spectrum that, using the mass and momentum conservation argument discussed above, must go as $P_{\omega \omega}\sim k^{4}$ and $P_{\omega \omega}\sim k^{7+2n}$, respectively. Because of the self-similarity of the system, the $k$-scaling is related to to the time dependence, which is found to be

\be\label{res1}
P_{\omega\omega}(k,a)\propto \left\{ \begin{array}{l} a^{(11-n)/(3+n)}\,k^{4} \\ a^{(17+3n)/(3+n)} \, k^{7+2n} \end{array} \right. \,,
\ee
For $n=-1.5$, relevant for our cosmology, both contributions are the same and one finds $P_{\omega\omega}\sim a^{8.3}$. In $\Lambda$CDM self-similarity is only broken by the non-power-law initial conditions and, at very late times $z\gtrsim 1$, by dark energy. It is argued that the latter is a small corrections so that the above results should be valid also in $\Lambda$CDM for $z\gtrsim 1$ with $n$ being the scalar spectral tilt around the non-linear scale. This remarkable result shows that \textit{the vorticity power spectrum at leading order is  determined by the symmetries of the problem and the power spectrum around the non-linear scale} (the cross correlations of $\bm{\omega}$ with $\theta$ and $\delta$ are trivially zero because of statistical isotropy and $\bm{\omega}$ being transverse).

\paragraph{Renormalization of the continuity equation} We show that \textit{the counterterms in the current formulation of the EFTofLSS} (see e.g.~\cite{Baumann2012,Carrasco2012,Hertzberg2012,Pajer2013}) \textit{are not sufficient to cancel all the divergences arising in the three correlators of density $\delta$ and velocity divergence $\theta$}. In order to renormalize the theory for $\delta$ and $\theta$, additional terms have to be added to the continuity equation. More specifically, at leading order one finds

\begin{equation}\label{contint}
\partial_{\tau}\delta + \partial_i  \left[ (1+\delta)  v^i \right] \eq -\chi_{1} \frac{\bigtriangleup\delta}{\cH} + \chi_{2} \frac{ \bigtriangleup\theta}{ \cH^{2}}-\frac{\tJ}{\cH}+\dots \;,
\end{equation}
where $\tJ$ is a new stochastic term, $\chi_{1,2}$ represent heat conduction and the ellipsis stands for terms that are of higher order in derivatives and perturbations. We show that these new terms as well as those already present in previous discussion of the EFTofLSS generically arise from the (perturbative) renormalization group flow and that they must be present since they are not forbidden by any symmetry of the problem. 
Note well that these new terms are only necessary to cancel the divergences that arise in the velocity correlators. The finite, cutoff independent part of $\chi_{1,2}$ and $\tJ$ can be chosen arbitrarily, for example it can be set to zero, by properly defining what we mean by $\vv$. We discuss how this provides a prescription to extract the velocity field from simulations or observations.

\paragraph{Renormalized correlators} We derive the renormalized two-point correlators for $\delta\delta$, $\delta\theta/\cH$ and $\theta\theta/\cH^{2}$ in an EdS universe with power-law initial conditions (a self-similar cosmology). In passing, we explicitly shown that only when the counterterms $\tJ$ and $\chi_{1,2}$ are included in the equations of motion all divergences in these correlators can be canceled. Finally, we find that the three correlators \textit{all share the same structure}
\begin{equation}\label{eq generalIintro}
\begin{split}
\Delta^{2}(k)\equiv \frac{k^{3}}{2\pi^2} P(k) \eq & \left( \frac{k}{\kNL} \right)^{n+3} \left\{ 1 + \left( \frac{k}{\kNL} \right)^{n+3} \left[ \alpha_1 + \tilde{\alpha}_1 \ln \frac{k}{\kNL} \right] \right\}   \\[1.5ex]
& + \beta \, \left( \frac{k}{\kNL} \right)^{n+5} + \gamma \, \left( \frac{k}{\kNL} \right)^{7} + \dots \;,
\end{split}
\end{equation}
where the time-dependent non-linear scale $k_{NL}(a)$ is defined by $\Delta_{lin}(k_{NL},a)=1$ and higher order terms in $k/k_{NL}$ are implied by the dots. We compute the coefficients $\alpha_1$ and $\tilde \alpha_1$ (different for each of the three possible correlators) using dimensional regularization and collect them in Tab.~\ref{tab alpha}. The six fitting parameters $\beta$ and $\gamma$ are determined by the five parameters $d^{2}$, $\dst$, $\ex{JJ}$, $\ex{J,\tJ}$ and $\ex{\tJ\tJ}$. Three of these can be fixed, e.g.~to zero, by appropriately defining the velocity. The remaining two should be fit to simulations or observations. Notice though that, since in our analysis we neglected two-loop corrections (an effort to include them in the EFTofLSS started in \cite{Carrasco2013}), only terms with dimension smaller than $(3+n)2$ should be included in the fit for consistency. In other words,  $\gamma$ should not be included for $n\leq -2/3$.

\paragraph{Momentum} The momentum is related to velocity and density by $\vp=\bar\rho(1+\delta)\vv +\textit{counterterms}$. The continuity equation becomes linear when the theory is formulated in terms of $\vp$
\be
\partial_{\tau}\delta+\nabla \cdot \vp=0\,.
\ee
Therefore, when using momentum instead of velocity, the only EFT corrections are in the effective stress tensor as discussed in the previous literature. This confirms the fact that the finite part of the counterterms needed to renormalize the velocity correlators only affect the matching relation to the momentum, i.e.~the definition of $\vv$, but not the number of free physical parameters in the theory. Although $\vp$ is a useful variable to use, e.g.~because it makes the continuity equation trivial and it satisfies the same $k^{4}$-law as $\vv$ (discussed above), it has a few drawbacks. First, as it is well-known, the vorticity $\vn\equiv \nabla \times \vp/\bar{\rho}$ of the momentum does not decouple from the equations for the density $\delta$ and divergence of the momentum $\mu\equiv \nabla \cdot \vp/\bar{\rho}$, thereby making the perturbation theory more cumbersome. Second, since neither $\vn$ nor $\mu$ are Galilean invariant, their correlators generically have \textit{uncancelled IR divergences} even for not so red spectra, namely $P_{\delta\delta,lin}\sim k^{n}$ with $n\leq-1$. This is to be contrasted with the Galilean invariant quantities $\theta$ and $\bm{\omega}$, for which IR divergences cancel as long as $n>-3$. 

\vspace{4ex}

Let us conclude with a few comments concerning the literature. The limits of the standard approach to cosmological perturbation theory have been addressed in Refs.~\cite{Pueblas2009,Valageas2011}, where the non-perturbative effect of shell-crossing has been discussed. Several issues that are addressed in the EFTofLSS, such as the divergences, the smoothing, etc., were not discussed. The authors of Ref.~\cite{Pietroni2012} considered the effects of smoothing out short scales. However, they considered the equation of motion of the stress tensor itself rather than making an Ansatz for it. Note that the renormalization procedure of the EFTofLSS, where we renormalize the coefficients in order to cancel the UV divergences, has little to do with the resummation schemes proposed e.g.~in Refs.~\cite{Crocce2006a,Crocce2006,Matarrese2007}. There, one usually tries include some, but not all, contributions from higher orders (see also Ref.~\cite{Sugiyama2013} for a recent discussion of the various approaches).

\paragraph{Note added} We are in debt to M.~Zaldarriaga for pointing out a crucial mistake in the first version of this manuscript, which lead us to the wrong conclusion about the renomalizability of the theory for $\delta$ and $\theta$. The current discussion of the renormalization of $\theta$ is now different from the previous version, while our results for the renormalized power spectra as well as for the vorticity remain unchanged.



\section{The Effective Field Theory of Large Scale Structures}\label{s:1}

The original Refs.~\cite{Baumann2012,Carrasco2012,Hertzberg2012} contain a detailed discussion of the construction of the EFTofLSS. In this section, we rederive the basic equations of the EFTofLSS and argue that terms which have been neglected in the literature must be included when considering the renormalization of the velocity. Throughout this paper we will assume an Einstein-de Sitter (EdS) universe, i.e. $\Omega_m = 1$. Therefore, the FLRW scale factor and the Hubble parameter take the simple form

\begin{equation}
\hubble \eq \frac{2}{\tau} \;, \qquad \qquad a \eq \frac{\tau^2}{\tau_0^2} \;, 
\end{equation}
where $\tau$ denotes conformal time.

Before proceeding, we have to discuss the basic degrees of freedom with which we will work. Besides the density contrast $\delta$, there are two natural choices: the velocity $\vv$ the momentum $\vp$. We will present results for both, starting with the velocity in section Sec.~\ref{s:2pt} and continuing with the momentum in Sec.~\ref{s:mom}. Some comments on the advantages and disadvantages of each choice are in order. The scalar part of the momentum, $\mu\equiv \nabla \cdot \vp/\bar \rho$ is \textit{linearly} related to the time derivative of the density by the continuity equation, hence it can be solved for at all orders. This cannot be done for the scalar component of the velocity, $\theta\equiv \nabla \cdot \vv$, which is non-linearly related to the density. Because of this non-linear relation, when using $\theta$ additional operators appear in the continuity equation from integrating out short modes as we will discuss in the next two subsections. Another argument in favor of using the momentum is the fact that the momentum, like the density, is straightforward to extract from numerical simulations, while, for the velocity, additional care is required as we discuss in Sec.~\ref{s:sims}. On the other hand there are two advantages in using the velocity. First, the vector part of the velocity, namely the vorticity $\vw\equiv \nabla \times \vv$, is very small in the perturbative regime and its effect on $\theta$ and $\delta$ can safely be neglected (see Sec.~\ref{s:vorticity} for a more detailed discussion). On the contrary, the vector part of the momentum, namely $\vn\equiv \nabla \times \vp/\bar \rho$, is of the same order as $\vm$ and cannot be neglected. Because of this, the use of velocity allows for a simpler form of the perturbative expansion. Furthermore, the correlators involving the velocity are free of IR divergences while momentum correlators are not, as we shall see Sec.~\ref{s:mom}. In any case, the two alternative description are related in a simple way as we will discuss in Sec.~\ref{ss:rel}.


\subsection{Symmetries}\label{ss:sym}

We are interested in describing a system of non-relativistic and collisionless point particles in Newtonian cosmology. Our goal is to derive the temporal evolution of some ensemble averages over stochastic initial conditions of the matter distribution. It has been shown that such a system can be described as a fluid. Hence, we assume that the system can be described using the number density $\rho$, with an average $\bar\rho$ and perturbations $\delta\equiv\rho/\bar \rho-1$, as well as a velocity field $\vv$ (see Sec.~\ref{s:mwv} and \ref{s:mom} for details on the definition of the velocity field). We start from the most generic form of the continuity and Euler equation, i.e.~containing at most one time derivative acting on the density contrast and ther velocity, and show how symmetries constrain it to take a much simpler from. For a Newtonian theory of dark matter particles the relevant symmetries are

\begin{enumerate}
\item Statistical homogeneity and isotropy \label{1}
\item Galilean invariance \label{2}
\item Conservation of the total number of dark matter particles \label{3}
\item Conservation of the total momentum \label{4}
\item Equivalence principle\label{5}
\end{enumerate}  

A few comments are in order. Statistical homogeneity and isotropy here means two things: first, all expectation values $\ex{\dots}$ can depend only on the norm of the distance between the points where the operators in the expectation value are evaluated; second, all numerical coefficients, as opposed to random variables, cannot depend on space, but are allowed to be arbitrary functions of time. Galilean invariance can be straightforwardly discussed in an expanding universe, but the additional factors of the scale factor $a$ and the Hubble parameter $\cH\equiv \partial_{\tau}a/a$ clutter the equations making it harder to follow the algebra. Since flat space contains already all the conceptual problem we want to discuss, we first restrict ourselves to this case. At the end we will rewrite our flat-space result in a way that is appropriate for an expanding universe.

Galilean invariance implies that the equations should take the same form under the transformation $\vx\rightarrow \vx+\vu t$ for some velocity $\vu$ with $t$ left invariant. The observables transform according to $\rho\rightarrow\rho$ and $\vv\rightarrow \vv-\vu$ (as well as $\vp\rightarrow \vp-\vu \rho$ for the momentum $\vp$). Time derivatives are not Galilean invariant by themselves, since $\partial_{t}\rightarrow \partial_{t}+\vu\cdot \nabla$ but should appear in the Galilean invariant combination $\partial_{t}+\vv\cdot \nabla$ (the convective derivative). The conservation of the total number of dark matter particles, which is a good assumption in most models of dark matter where interactions and decay rates are very small, implies that\footnote{This is a Galilean invariant definition because of the usual assumption that fields vanish at large distances so that the boundary terms of space integral can be dropped.} $\int d^3x \, \partial_{t}\rho=0$. The conservation of momentum 
leads to $\int d^3x\, \partial_{t} (\vv\rho)=0$. Finally the equivalence principle guarantees that all bodies in the same gravitational field accelerate in the same way. This can be though of a definition of what we mean by the gravitational potential $\phi$.

 
\subsection{The equations of motion}\label{ss:eom}

Now that we have spelled out the symmetries in detail we are ready to use them to constrain the form of the equations of motions. We start considering generalized continuity and Euler equations which contain $\partial_{\tau}\rho$ and $\partial_{\tau} \vv$, respectively. Since we can multiply each equation by an arbitrary non-vanishing factor, we can fix to one the coefficient of these time derivatives. It is convenient to introduce the divergence and the vorticity of the velocity by

\begin{equation} \label{eq vdecomp}
\theta \eq \vec{\nabla} \cdot \vec{v} \;, \qquad \qquad \bm{\omega} \eq \vec{\nabla} \times \vec{v} \;.
\end{equation}
Since the vorticity decays in linear theory, its effect on $\delta$ and $\theta$ are negligible as we shall discuss more carefully in Sec.~\ref{s:vorticity}. For this section we therefore neglect $\vw$ and come back to it in Sec.~\ref{s:2pt}. Arranging the terms in Galilean-invariant combinations, we hence write

\begin{equation}\label{eq ansatz}
\begin{split}
& \left(\partial_{\tau}+\vv\cdot \nabla \right)\rho+ C_{\theta} \theta+C_{\rho \theta} \,\rho \theta \eq -\tJ+C_{\rho} \rho + C_{\bigtriangleup\delta}\bigtriangleup\delta+C_{\bigtriangleup\theta}\bigtriangleup\theta +\dots\,,\\[1.5ex]
& \left(\partial_{\tau}+\vv\cdot \nabla \right)\vv+C_{ \nabla \phi} \nabla \phi \eq  -{\bf J}+C_{\nabla\rho}\nabla \rho + C_{\nabla \theta}\nabla \theta + C_{(\rho+\theta)}(\rho+\theta)\nabla \phi+\dots\,,
\end{split}
\end{equation}
where the ellipsis stands for terms that are of higher order in the derivative expansion or in power of the perturbations $\delta$ and $\vv$. Because of symmetry \ref{1}, all numerical coefficients $C$ are at most functions of time. We have also included two noise terms $\tJ$ and $J$, whose sign is pure convention, that are stochastic variables rather than numbers, which means that they depend on space and time provided their ensemble averages respect symmetry \ref{1}. As we discuss in detail in Appx.~\ref{sec JJ} symmetries \ref{3} and \ref{4} constrain ${\bf J}$ and $\tJ$ to start at order $k^{2}$ in Fourier space. Notice that $\vv$ cannot appear without derivatives (unless in the appropriate combination with $\partial_{\tau}$ which we already included) because of symmetry \ref{2} and $\phi$ without derivatives is unphysical and is not included. Symmetry \ref{3} fixes $C_{\rho}=0$ and $C_{\rho\theta}=1$ while symmetry \ref{4} leads to $C_{\theta}=1$. Finally, symmetry \ref{5} implies $C_{(\rho+\theta)}=0$ and $C_{\nabla \phi}=1$, which means that the field $\phi$ that appears in Eq.~\eqref{eq ansatz} is the same as the one in the Poisson equation. Notice that, by peeking at the final solution, we have arranged all terms with unity coefficients on the left-hand side while all the forbidden and unconstrained terms are on the right-hand side.

At this order in derivatives and powers of the perturbations, we can rewrite the above equations of motion using more physically transparent names for some of the parameters. Also, it is now straightforward to include the terms related to the expansion of the universe. The result, including the Poisson equation, is

\be
\partial_{\tau}\delta + \partial_i  \left[ (1+\delta)  v^i \right] &\eq& -\chi_{1} \frac{\bigtriangleup\delta}{\cH} + \chi_{2} \frac{ \bigtriangleup\theta}{ \cH^{2}}-\frac{\tJ}{\cH} \;, \label{ce}\\
 \bigtriangleup \phi  &\eq& \frac{3}{2} \hubble^2 \, \delta \;, \label{Pe}\\
 \dot{v}^i + \hubble \, v^i + \partial^i \phi + v^j \, \partial_j v^i  &\eq&  - c_s^2 \, \partial^i \delta + \frac{3}{4} \frac{c_{sv}^2}{\hubble} \, \bigtriangleup v^i + \frac{4c_{bv}^2 + c_{sv}^2}{4 \hubble} \, \partial^i \vec{\nabla} \cdot \vec{v} - \Delta J^i \;.\label{Ee}
\ee
We have included an appropriate factor of $\cH^{-1}$ so that $\tJ$ and $J\equiv \nabla_{i} \cdot \Delta J^{i}$ have both dimension of a mass square and $\chi_{1,2}$ and $c_{s,sv,bv}$ are dimensionless. Although $J$, $\tJ$, $\chi_{1,2}$ and $c_{s,sv,bv}$ are functions of time, we shall adopt the EFT jargon and call them low-energy constants (LECs). As usual in EFT they are not determined by the theory itself and have to be extracted form simulations or observations. Note well that the continuity equation contains additional terms compared to the equations of motion discussed in Refs.~\cite{Baumann2012,Carrasco2012,Hertzberg2012,Pajer2013}. Let us give three arguments for the necessity of the terms proportional to $\chi_{1,2}$ and $\tilde{J}$. The first reason is that, as explained in this section, these additional terms are not forbidden by any symmetry and therefore should be included following the EFT approach. The second reason is that these terms are needed to cancel the divergences arising in the correlators for $\delta$ and  $\theta$. In Ref.~\cite{Pajer2013} it was shown that the terms in the right hand side of \eqref{Ee} are sufficient to cancel all divergences in the $\delta$ power spectrum at one loop.
As we will see, this ceases to be true if one wants to compute correlators including $\theta$, for which the terms in the right-hand side of \eqref{ce} are needed. The third reason is that one can see these terms arise when following the renormalization group flow in the perturbative regime, as we explain in the next subsection.

Before proceeding let us give some physical interpretation for the new corrections to the continuity equation. The upshot will be that, although they are needed to cancel divergences arising from loop corrections, their finite, cutoff-independent part can be chosen arbitrarily, reflecting the freedom in defining the velocity field. Let us see this in more details (e.g.~following the discussion in Appx.~B.10 of \cite{Weinberg:2008zzc}). A perfect fluid is defined by the property that in the rest frame of any given fluid element the energy-momentum tensor $T^{\mu\nu}$ is diagonal and isotropic and, if a conserved charge $n$ is present, in the rest frame it is described by $N^{\mu}_{;\mu}=\dot n=0$ (neglecting the expansion of the universe). Our dark matter fluid is a non-relativistic system and therefore the conservation of energy coincides with that of dark matter particle number, so we will not discuss it further. For an imperfect fluid the situation changes because, unlike for a perfect fluid, there is more than one velocity. There is the velocity of the transport of energy and that of momentum, which are in general different from each other. In principle any other velocity, different from these two can be used. Because of this, there is no reference frame in which $T^{\mu\nu}$ is diagonal and isotropic (and the number conservation is $\dot n=0$). For a fluid that is not too far from a perfect one, this fact can be captured by adding higher derivative corrections to $T^{\mu\nu}$ (and $N^{\mu}$). The corrections induce dissipation and lead to the production of entropy (which is conserved for a perfect fluid). All the EFT corrections we found in this section are a specific example of this discussion. There is an important distinction though between the corrections in the continuity and Euler equation, which we will discuss in more detail in Sec.~\ref{s:sims}. The finite part of the corrections in the continuity equation can be chosen arbitrarily by changing our definition of $\vv$. So for example, although the infinite, cutoff dependent, unphysical part of these terms is needed in order to explicitly cancel the cutoff dependence of loop corrections (the ``UV-divergences'', which are not necessarily divergent but always unphysical), the finite part can be set to zero, or any other convenient value. In the following, we will derive all formulae for the general case, i.e.~with non-zero $\chi_{1,2\,\textit{ren}}$ and $\tilde{\vec{J}}_\textit{ren}$, but one should bear in mind that the only physical parameters that have to be fitted either to simulations or data are those in the Euler equation.

 
\subsection{Integrating out short modes}\label{ss:out}

The reason why all operators compatible with the symmetries of the problem must be included in an EFT is particularly clear following the Wilsonian approach. Consider a theory defined below a certain cutoff scale $\Lambda$. One can derive the same theory defined with a lower cutoff $\Lambda -d\Lambda$ by integrating out all modes with momentum between $\Lambda$ and $\Lambda-d\Lambda$. In an ideal world we would like to do this with all the short scale modes in LSS, but since these evolve in a complicated non-linear way, this is extremely hard to do. On the other hand, what we can do is to perform the computation in perturbation theory. In general, the new terms generated by this perturbative computation must be a subset of all possible terms generated by performing the fully non-linear computation. In practice, since the perturbative theory does not obey any additional symmetry as compared with the non-linear theory, all operators are found (at least at the 1-loop order we are considering here). Notice that a similar computation was presented in Ref.~\cite{Baumann2012}.

Let us start with the standard SPT equations, forgetting for the moment the EFT corrections in the right-hand sides of \eqref{ce} and \eqref{Ee}. Neglecting for simplicity the vorticity one finds

\begin{equation}\label{noEFT}
\begin{split}
\partial_{\tau} \delta+ \theta \eq &- \intq \alpha(\vec q,\vec k-\vec q) \theta(\vec q)  \delta(\vec k-\vec q)\,,\\[1.5ex]
\partial_{\tau} \theta+ \cH \theta+\frac{3}{2} \cH^{2} \delta \eq & -\intq \beta(\vec q,\vec k-\vec q) \theta(\vec q)\theta(\vec k-\vec q) 
\end{split}
\end{equation}
where 
\be\label{conv}
\intq\equiv \int \frac{d^{3}q}{(2\pi)^{3}}\,,\quad\alpha (\vec q_{1},\vec q_{2})\equiv\frac{(\vec q_{1}+\vec q_{2})\cdot \vec q_{1}}{q_{1}^{2}} \,, \qquad \beta(\vec q_{1},\vec q_{2})\equiv \frac{ \left(\vec q_{1} +\vec q_{2}\right)^{2}\vec q_{1} \cdot \vec q_{2} }{2 q_{1}^{2}q_{2}^{2}}\,.
\ee
The perturbative solutions of these equations can be calculated following SPT. As described in detail in \cite{Bernardeau2002}, the density contrast and the velocity dispersion are Fourier transformed and expanded\footnote{Notice that our notation differs from that in \cite{Bernardeau2002} in two regards. First, our Fourier convention is as in \eqref{conv} while in \cite{Bernardeau2002} there no $(2\pi)^{3}$ in $d^{3}q$. Second, we re-absorb the $a^{n}$ into $\delta_{n}$ and $\theta_{n}$.} in a series of increasing powers of $a$ 

\begin{equation}\label{eq SPTdecomp}
\delta(\vec{k},\tau) \eq \sum_n \delta_n (\vec{k}) \;, \quad  \theta(\vec{k},\tau) \eq \sum_n \theta_n (\vec{k})\quad \mathrm{with}\quad  \theta_{n}/\hubble\propto\delta_{n}\propto a^{n} \;.
\end{equation}
The $n^\mathrm{th}$ order contribution involves $n$ powers of $\delta_1(\vec{k})$ and it is given by

\begin{equation}\label{eq loop}
\begin{split}
& \delta_n(\vec{k}) \eq \int_{\vk_{1},\vk_{2},\dots,\vk_{n}} \;(2\pi)^{3} \dirac(\vec{k} - \sum_{i=1}^n \vec{k}_i ) \, F_n(\vec{k}_1, \dots , \vec{k}_n) \, \delta_1(\vec{k}_1) \cdots \delta_1(\vec{k}_n) \;, \\[1.5ex]
&\theta_n(\vec{k}) \eq  -\hubble\int_{\vk_{1},\vk_{2},\dots,\vk_{n}}\; (2\pi)^{3}\dirac(\vec{k} - \sum_{i=1}^n \vec{k}_i ) \, G_n(\vec{k}_1, \dots , \vec{k}_n) \, \delta_1(\vec{k}_1) \cdots \delta_1(\vec{k}_n) \;,
\end{split}
\end{equation}
where the kernels $F_n$ and $G_n$ are dimensionless functions of momenta ($F_n^s$ and $G_n^s$ denote the symmetrized kernels) for which recursive relations are available\footnote{Notice that our convention is different from the one in \cite{Bernardeau2002} in two aspects: we absorb the time dependence of $\delta$ in $\delta_n$ and $\int_\vq = \int d^3q / (2\pi)^{3}$.}. 

Let us see how this machinery leads to additional terms once the short modes of $\delta$ and $\theta$ are integrated out in the case of the interaction proportional to $\alpha$. 
At second order in perturbations the $\alpha$ schematically interaction reads $\int \alpha \delta_{1}\theta_{1}$. Each of the first order perturbation can be either a short mode, with momentum between $\Lambda$ and $\Lambda-d\Lambda$, or a long mode, with momentum below $\Lambda-d\Lambda$. We want to integrate out the former and see their effect on the latter. Dropping the index $1$ in order to simplify the notation, we can write

\be\label{sto}
\intq \alpha(\vec q,\vec k-\vec q)  \left[ \delta_{l}(\vec q) \theta_{l}(\vec k-\vec q)+\delta_{s}(\vec q) \theta_{s}(\vec k-\vec q)\right] \,.
\ee
Notice that one cannot construct a long wavelength perturbation with only a long and a short mode. The first term is the same interaction as we had before, which does not get renormalized at this order. The second term is a stochastic source analogous to ${\bf J}$ and $\tJ$ in the previous section. This shows that this operator is generated already in perturbation theory by integrating out short modes and therefore cannot not be omitted.

We can go one more step and expand the interaction $\int \alpha\theta \delta$ at cubic order as
\be\label{al}
&&\intq \alpha(\vec q,\vec k-\vec q)  \left[ \theta_{2}(\vec q)  \delta_{1}(\vec k-\vec q)+\theta_{1}(\vec q)  \delta_{2}(\vec k-\vec q)\right]\\
&=&\intq \alpha(\vec q,\vec k-\vec q)  \left[ \intqp G_{2}^{s}(\vec q',\vec q-\vec q') \delta_{1}(\vec q')\delta_{1}(\vec q-\vec q')  \delta_{1}(\vec k-\vec q)+ \right.\nonumber\\
&&\left.\hspace{1cm}\quad\intqp F_{2}^{s}(\vec q',\vec k-\vec q-\vec q') \delta_{1}(\vec q)  \delta_{1}(\vec k-\vec q-\vec q')\delta_{1}(\vec q')\right]\,.
\ee
As before each of the first order perturbations can be a short or a long mode. Three long modes give us back the interaction we started with. Having only one short and two long modes cannot lead to a long wavelength perturbation. If all three perturbations are short we get again a stochastic term as from \eqref{sto}. A new term instead arises when two perturbations are short and one is long. To find the net effect on long modes we take the expectation value over short modes. Fluctuations on top of this expectation value are indeed present but turn out to be subleading corrections in the final power spectrum. There are several options for contracting two short modes. It is easy to realize that contracting the perturbations $\ex{\delta_{s}(\vec q')\delta_{s}(\vec q-\vec q')} $ and $ \ex{\delta_{s}(\vec q) \delta_{s}(\vec q')}$ in \eqref{al} gives zero because the kernels vanish. The other two contractions for each term, namely $\ex{\delta_{s}(\vec q-\vec q')  \delta_{s}(\vec k-\vec q)}$ and $\ex{\delta_{s}(\vec q')\delta_{s}(\vec k-\vec q)}$ for the first and $\ex{\delta_{s}(\vec q)  \delta_{s}(\vec k-\vec q-\vec q')}$ and $\ex{\delta_{s}(\vec q')  \delta_{s}(\vec k-\vec q-\vec q')}$ for the second, give the same contribution. Using the delta function from these short-mode contractions to perform the $\vec q'$ integral, we end up with
\be
\intq \alpha\, \theta \delta&\supset&\delta(\vec{k}) \intq\,2\, \alpha(\vec q,\vec k-\vec q)  \left[  G_{2}^{s}(\vec k,\vec q-\vec k) P(q)+ F_{2}^{s}(\vec k,-\vec q) P(q)\right]\,.
\ee
The leading term in an expansion of $k\ll q$ is indeed of the form $C_{\bigtriangleup \delta} \bigtriangleup\delta$ with higher order terms representing higher derivative corrections. The precise value of the coefficient $C_{\bigtriangleup\delta}$ in perturbation theory can be different from the full non-perturbative result if the modes we are integrating out are well into the non-linear regime. But the structure of the term, i.e.~$\bigtriangleup\delta$, is robust. This shows that these operators are generated by the renormalization group and should not be neglected. Notice that at this order $\delta$ and $\theta$ are indistinguishable since $\delta_{1}=\theta_{1} $. This means that the argument above is valid also to justify $\bigtriangleup\theta$. 

In summary, already integrating out short modes in perturbation theory shows the presence of additional terms in the continuity equation (as well as in the Euler equation as it has already been noticed in the original formulation of the EFTofLSS). The relation of these terms to the renormalization of composite operators will be discussed in section \ref{s:mom}. 

\subsection{EFT corrections in perturbation theory}

Having found the equations \eqref{ce}, \eqref{Pe} and \eqref{Ee}, we can now proceed to solve the theory perturbatively. We will have two expansions. The first expansion is in powers of the perturbations as in SPT. The perturbations at various orders are then given by \eqref{eq loop}. The second expansion is in powers of the EFT corrections. We will consider here only corrections to the perturbations at linear order in $\tJ$, ${\bf J}$, $c_{s,sv,bv}^{2}$ and $\chi_{1,2}$, which turns out to be sufficient to cancel all divergences at one-loop order. Also, at lowest order we can simplify the EFT terms
\be
\partial_{i}\left(- c_s^2 \, \partial^i \delta + \frac{3}{4} \frac{c_{sv}^2}{\hubble} \, \bigtriangleup v^i + \frac{4c_{bv}^2 + c_{sv}^2}{4 \hubble} \, \partial^i \vec{\nabla} \cdot \vec{v} \right) &=& -\dss \bigtriangleup\delta+\mathcal{O}(d_{s}^{4})\,,\\
 -\chi_{1} \frac{\bigtriangleup\delta}{\cH} + \chi_{2} \frac{ \bigtriangleup\theta}{ \cH^{2}} & =& -\dst \frac{\bigtriangleup\delta}{\cH}+\mathcal{O}(\tilde d_{s}^{4})
\ee
where we have defined
\be
\dss\equiv c_{s}^{2}+c_{sv}^{2}+c_{bv}^{2}\,,\quad \dst\equiv\chi_{1}+\chi_{2}\,,
\ee
and used that $a\partial_{a}\delta_1 = \delta_1 = -\theta_1/\cH$.

The full set of non-linear equations for $\delta$, $\theta$ and $\bm{\omega}$ is found in Appx.~\ref{sec fullEOM}. The linearized equations of motion for $\delta$, $\theta$ and $\bm{\omega}$ at first order in the LECs are given by

\be\label{eq eomlin}
  \hubble^2 \, \left\{ -a^2 \partial_a^2 - \frac{3}{2}  a \partial_a + \frac{3}{2 } \right\} \delta &\eq &-d^{2}\bigtriangleup  \delta_{1}- J+\left(\frac{3}{2}+a\partial_{a}\right) \left(\dst \bigtriangleup\delta_{1}+\tJ \right) \;, \\[1.5ex]
 \label{eq eomlin2} \hubble \left\{ a^2 \partial_a^2 + \frac{5}{2} a \partial_a - 1 \right\} \theta & \eq &  - \left(1+a\partial_{a}\right) \left( \dss \bigtriangleup \delta_{1}+J\right)+\frac{3}{2}\left(\dst \bigtriangleup\delta_{1}+\tJ \right) \,, \\[1.5ex]
\label{eq eomlin3} \hubble \, \Big\{  a \partial_a  + 1 \Big\} \, \bm{\omega} &\eq & \frac{3}{4} \frac{c_{sv}^2}{\hubble} \bigtriangleup \bm{\omega} - \vec{\nabla} \times \Delta \vec{J} \;.
\ee
The retarded Green's function that follow from the homogeneous part of the above equations can be found by solving \eqref{eq eomlin}, \eqref{eq eomlin2} and \eqref{eq eomlin3} at time $a$ with the right-hand side substituted with $+\delta_{D}\left(a-a'\right)$. 
One finds 
\begin{equation}\label{eq green}
\begin{split}
G_\delta (a , a')  \eq & \theta_H(a-a')  \, \frac{2}{5 \, \hubble_0^{2}} \left\{ \left( \frac{a'}{a} \right)^{3/2} - \frac{a}{a'}  \right\} \\[1.5ex]
G_\theta (a, a') \eq & - \frac{\hubble_0}{\sqrt{a}} \, G_\delta (a, a') \;, \\[1.5ex]
G_\omega (a, a') \eq & \theta_H(a-a')  \,  \frac{\sqrt{a'}}{a} \frac{1}{\hubble_0} \;, \\[1.5ex]
\end{split}
\end{equation}
where $\hubble_0 = 2/\tau_0 = \hubble \sqrt{a}$. Setting the LECs to zero, the standard solution for $\delta$ with a growing mode proportional to $a$ and a decaying mode proportional to $a^{-3/2}$ is recovered. Also, the vorticity decays in linear theory as $\vw\sim a^{-1}$, as expected.

For the moment, let us neglect the vorticity, which will be discussed at in Sec.~\ref{s:mwv} and \ref{s:vorticity}. We denote corrections to the linear solutions of Eqs.~\eqref{ce} and \eqref{Ee} that are induced by the EFT terms as $\delta_c$ and $\delta_J$ (and analogously for $\theta$) and compute them using the Green's functions above. To perform the time integral and evaluate the time derivatives $a\partial_{a}$ we need to know the time dependence of the LECs. This can be inferred as follows. The LECs contain a cutoff dependent counterterm\footnote{We will use both cutoff and dimensional regularization whichever is more useful. For general discussions, we think that the cutoff regularization allows for a more intuitive understanding.}, which will cancel the appropriate loop divergence, as well as a finite, cutoff independent renormalized part, i.e.

\begin{equation}
d^2 \eq d_\textit{ren}^2 + d_\textit{ctr}^2 \;, \quad \qquad J \eq J_\textit{ren} + J_\textit{ctr} \;,
\end{equation}
and likewise for $\tilde{d}$ and $\tilde{J}$. The time dependence of the counterterm must be the same as that of the loop divergence for a cancellation to be possible, then one must have schematically
\be
\ex{\delta_{J}\delta_{J}}&\sim &\int da' G_{\delta}(a,a') \int da'' G_{\delta}(a,a'') \ex{J(a')J(a'')} \sim \ex{\delta_{2}\delta_{2}}\sim a^{4}\,,\\[1.5ex]
\ex{\delta_{c}\delta_1}&\sim &\int da' G_{\delta}(a,a') \dss \ex{\delta_{1}(a')\delta_{1}(a)} \sim \ex{\delta_{1}\delta_{3}}\sim a^{4}\,.
\ee
Notice that $\int da' G_{\delta}(a,a')\sim a$ and in fact one finds the useful relation
\be
\int da' \, G_{\delta}(a,a')\,a'^{p}=-\frac{a^{p+1}}{\cH_{0}^{2} \,p(p+5/2)}\,.
\ee
We therefore deduce
\be\label{vio}
d_{ctr}^{2}\sim \tilde d_{ctr}^{2} \propto  a\,,\quad\quad J_{ctr}\sim \tJ_{ctr}\propto a\,,
\ee

Let us turn to the finite and cutoff independent part of the LECs. In general it would be unconstrained and one would need to extract the time-dependent renormalized value from observations or simulations\footnote{This is true once one has provided a specific definition of $\vv$, such as a procedure to extract if from the data, which is what we are assuming in this discussion. As we will discuss in Sec.~\ref{s:mom}, one can actually fix the value of $\tilde{d}^2_\textit{ren}$ and $\tJ_\textit{ren}$ to \textit{define} what is meant by $\vv$.}. On the other hand, as we will study in detail in Sec.~\ref{sec selfsim}, in an EdS universe with power-law initial conditions there is a symmetry relating the spatial and time dependence of all terms, known as \textit{self-similarity}. This symmetry will uniquely fix the time dependence of the renormalized LECs. Why is the time dependence of the counterterms not constrained by the same symmetries as the renormalized part? The reason is that cutoff renormalization breaks self-similarity since the cutoff introduces a fixed scale. Self-similarity is indeed recovered in the final, cutoff independent result, but the intermediate steps are not manifestly self-similar. The scaling in \eqref{vio} is precisely an example of this fact: cutoff dependent counterterms, being unphysical, have a scaling that violates self-similarity in such a way to precisely cancel the self-similarity-violating loop divergences. In this section we focus on the cutoff dependent part to show that all divergences cancel. The renormalized part will be discussed in Sec.~\ref{s:mom} and \ref{s:2pt}.

With this knowledge we can compute the time derivatives in \eqref{eq eomlin} and \eqref{eq eomlin2} as well as perform the time integral over the Green's function times the source

\be\label{eq deltacj}
 \delta_c (\vec{k}, \tau)& \supset&  \int da'\; G_\delta(a, a') \left[-\dss_{ctr}(\Lambda,a') \bigtriangleup \delta_{1}(a')+ \left(\frac{3}{2}+a'\partial_{a'}\right)\dst_{ctr}(\Lambda,a') \bigtriangleup \delta_{1}(a')\right] \nonumber \\
 &=&   \frac{ \, \bigtriangleup \delta_1}{\cH^{2}}\left[\frac{1}{9}\dss_{ctr}(\Lambda,a)-\frac{7}{18}\dst_{ctr}(\Lambda,a)\right] \;,\\[1.5ex]
 \theta_c (\vec{k}, \tau)& \supset&   \int da'\; G_\theta(a, a') \left[-\left(1+a'\partial_{a'}\right)\dss_{ctr}(\Lambda,a') \bigtriangleup \delta_{1}(a')+ \frac{3}{2}\dst_{ctr}(\Lambda,a') \bigtriangleup \delta_{1}(a')\right] \nonumber \\
 &=&   -\frac{ \, \bigtriangleup \delta_1}{\cH}\left[\frac{1}{3}\dss_{ctr}(\Lambda,a)-\frac{1}{6}\dst_{ctr}(\Lambda,a)\right] \;,\\[1.5ex]
\delta_J (\vec{k}, \tau)& \supset & \int da'\; G_\delta(a, a') \,\left[- J_{ctr}(\Lambda,a')+ \left(\frac{3}{2}+a'\partial_{a'}\right) \tJ_{ctr}(\Lambda,a')\right] \nonumber \\
&=& \frac{1}{\cH^{2}} \left[\frac{2}{7}J_{ctr}(\Lambda,a)-\frac{5}{7}\tJ_{ctr}(\Lambda,a)\right]\,,\\[1.5ex]
\theta_J (\vec{k}, \tau) &\supset & \int da'\; G_\theta(a, a') \,\left[-\left(1+a'\partial_{a'}\right)  J_{ctr}(\Lambda,a')+\frac{3}{2} \tJ_{ctr}(\Lambda,a')\right] \nonumber \\
&=&- \frac{1}{\cH} \left[\frac{4}{7}J_{ctr}(\Lambda,a)-\frac{3}{7}\tJ_{ctr}(\Lambda,a)\right]\,.
\ee

Similarly to $\delta$ and $\theta$, the EFT terms in Eq.~\eqref{eq eomlin3} generate two contributions to $\bm{\omega}$

\be
\bm{\omega}_c(a) &\eq &\int da'\; G_\omega(a, a') \frac{3}{4} c_{sv}^2(a') \, \frac{\bigtriangleup \vw(a')}{\cH(a')} \,,\\
 \qquad \bm{\omega}_J(\vec{k},\tau) &\eq& - \int da'\; G_\omega(a, a') \,  \vec{\nabla} \times \Delta \vec{J}  \;. 
\ee
%


\section{Consequences of mass and momentum conservation}\label{s:mwv}

Before we present results for the spectra of $\delta$, $\vec{v}$ and $\vp$, we want to clarify the relation between velocity and momentum. Then, in Sec.~\ref{ss:Pa}, we will present a general argument that puts constraints on the effect that short scales can have on long scales, more specifically on the stochastic noise. These constraints will be useful in determining the power spectrum of vorticity as we will see in Sec.~\ref{s:vorticity}

 
\subsection{Momentum and mass-weighted velocity}\label{ss:rel}

Momentum is straightforwardly defined by the continuity equation 
\be
\partial_{\tau} \delta+\mu&=&0\,, 
\ee
where we have defined $\mu\equiv \nabla \cdot \vp/\bar \rho$. This is a linear relation and therefore it is not modified when integrating out short modes. In other words it is stable under RG flow. Comparing this with the continuity equation written in terms of the velocity \eqref{ce}, one deduces the relation
\be \label{eq defpi}
\frac{\vp}{\bar{\rho}} \; \equiv \; (1 + \delta) \, \vv  + \frac{\chi_1}{\hubble} \, \vec{\nabla} \cdot \delta - \frac{\chi_2}{\hubble^2} \, \vec{\nabla} \cdot \theta + \frac{\tilde{\vec{J}} }{\hubble} +... \;,
\ee
where $\nabla\cdot \tilde{\vec{J}}=\tJ$ \footnote{In principle there should be also a term linear in the vorticity $\vw$, which we omitted in the previous section because we were and continue to neglect the vorticity.}. All fields appearing in this relation are smoothed and contain only long fluctuations (we are suppressing the label $l$ to simplify the notation). The reason why the standard relation $\vp=\bar\rho (1+\delta)\vv$ is corrected by an infinite series of operators organized in powers of perturbations and derivatives is the same as in Sec.~\ref{ss:eom}. Once short modes are integrated out, composite operators such as $\delta \vv$, i.e.~containing the product of operators at the same point, generate new terms that can be expanded in power of the long modes and their derivatives. The physical meaning of the EFT corrections appearing in \eqref{eq defpi} is best discussed after we have presented the results of the normalization process, so we defer it to Sec.~\ref{s:sims}.

Notice that the velocity defined in this way is a mass-weighted velocity, in the sense that it is given by $\smo{\bm \pi}/\smo{\rho}$ plus a prescription to handle this composite operator (namely adjusting the EFT coefficients in \eqref{eq defpi} to cancel UV-divergences). This is to be contrasted with the volume-weighted velocity, which has received some attention in the literature (see e.g.~\cite{Bernardeau1995,Bernardeau1996}). In this work and in the rest of the EFTofLSS literature, only the mass-weighted velocity is considered and therefore we differ some comments on the relation to volume-weighted velocity to Appx.~\ref{a:vwv}.

 
\subsection{Constraints on the large-scale effects of short scales}\label{ss:Pa}

It is well known that momentum and mass conservation put constraints on the large-scale corrections to density correlators that can be induced by short scales \cite{Peebles1980,Zeldovich1965}. The details of this and of the following computations are not too hard to follow, but are quite lengthy. Hence we prefer to collect them in Appx.~\ref{sec JJ} and only present the final results in the following. Let us start to consider some distribution of matter, separate large and short scales and concentrate on short scales. Short scales evolve according to complicated highly non-linear equations. Although we cannot solve these equations analytically, we can find some results in the large-scale limit $k\rightarrow 0$. Expanding $\delta_{s}(\vk)$ around $k=0$ one finds that the leading ($k^{0}$) and first subleading ($k^{1}$) terms are not arbitrary. Instead they are fixed once and for all by mass and momentum conservation. The first term in $\delta_{s}(\vk\rightarrow 0)$ that is not fixed by mass and momentum conservation but depends on the details of the short-scale dynamics comes at order $k^{2}$. This means that the corrections that can come from the short-scale dynamics to the density at large scales must decay quite fast. In formulae, we can usefully summarize this by 
\be\label{ok}
\ex{\delta_{s}(\vk)\delta_{s}(\vk)}\sim k^{4}
\ee
for $k\rightarrow 0$, plus higher order corrections. This result explains a couple of otherwise surprising facts about perturbation theory. Despite the fact that there are several terms in $P_{\delta\delta,22}(k\rightarrow 0)$ of order $k^{2}$, they all cancel each other out (the terms $k^{3}$ cancel out after performing the angular integral as expected on the ground of rotational invariance). Similarly, all the symmetrized SPT kernels $F_{n}^{s}$ and $G_{n}^{s}$ start at order $k^{2}$ in the large-scale limit.

The curious reader will now wonder how this discussion extends to the velocity. As we mentioned above, also the SPT kernels $G_{n}^{s}$ for $\theta$ start at order $k^{2}$ and so one finds $P_{\theta\theta,22}(k\rightarrow 0)\sim k^{4}$. Is there a conservation argument for $\theta$ analogous to the one for $\delta$? The answer is yes as we will now argue. First, conservation of momentum can affect in a simple way only the mass-weighted velocity and not the volume-weighted one, since only the former is directly related to momentum\footnote{For example, consider the momentum exchanged via the gravitational force between two particles. By Newton's third law the total momentum is unchanged and so must be the mass-weighted velocity which is just proportional to it. On the other hand, if one of the two particles is in an over-dense region, while the other in an under-dense region, the former must contribute more to the volume-weighted velocity (since there are less particles around). Then the result of the interaction will be an overall change (increase or decrease depends on which particle is faster) in the total volume-weighted velocity. }. 

In order to find some conservation equation for the mass-weighted velocity we start considering momentum. A derivation very similar to the one for the density (see Appx.~\ref{sec JJ}) gives $\vp\sim k^{1}$ and so the divergence $\mu$ and vorticity $\vn$ of the momentum have power spectra that satisfy
\be\label{mo}
\ex{\mu_{s}\mu_{s}}\sim \ex{\vn_{s}\cdot \vn_{s}}\sim k^{4}\,,
\ee
in the limit $k\rightarrow 0$ up to higher order corrections. In words, short-scale corrections to the large-scale power spectra of $\mu$ and $\bm{\nu}$ are suppressed by at least $k^{4}$ due to mass and momentum conservation. By generality arguments we expect this scaling to be saturated. As we mentioned earlier, the derivatives of the momentum are not Galilean-invariant quantities and so they exhibit uncancelled IR divergences. Therefore it is convenient to derive the consequences of Eq.~\eqref{mo} for $\vv\equiv \vp/\rho$. The complication here arises from dealing with the Fourier transform of the inverse. In Appx.~\ref{sec JJ} we expand $\vv$ for small $\delta$ and show that the non-linear terms $\vp \delta^{n}$ also obey the same conservation equation as $\vp$ if the masses of all particles are the same, as it is almost always the case for simulations, and in the limit in which the cutoff is removed\footnote{This is the most convenient limit to study the EFTofLSS because all terms suppressed by $\Lambda$ vanish.}$\Lambda\rightarrow \infty$. The consequences of having unequal masses require further consideration. 

Summarizing, we can show that both the divergence $\theta$ and the vorticity $\bm{\omega}$ of the velocity enjoy conservation equations analogous to the one for $\delta$, namely
\be\label{imp}
\ex{\theta_{s}\theta_{s}}\sim \ex{\bm{\omega}_{s} \cdot \bm{\omega}_{s}}\sim k^{4}\,,
\ee
in the limit $k\rightarrow 0$ up to higher order corrections. This result justifies the fact that also the SPT kernels for $\theta$ start at order $k^{2}$ or, equivalently, that $P_{\theta_{2}\theta_{2}}\sim k^{4}$. 

Eq.~\eqref{imp} has interesting consequences for the power spectrum of the vorticity $\bm{\omega}$, as we will discuss at length in Sec.~\ref{s:vorticity}. Since the linear solution of $\bm{\omega}$ decays, we expect that one of the leading contribution to $P_{\omega \omega}$ comes from the stochastic noise generated by short scales. The power spectrum of this noise must obey Eq.~\eqref{imp}. As we will discuss in the next section, if one then uses self-similarity, the $k^{4}$ scaling uniquely fixed the time dependence of $P_{\omega \omega}$ according to \eqref{eq Pwwgeneral} and \eqref{eq mnw}. As pointed out in \cite{Carrasco2013a} and further discussed in Sec.~\ref{s:vorticity}, another set of EFT corrections appear at a comparable order, depending on the value of $n$. Altogether this is neat result: the vorticity power spectrum is determined by the symmetries of the problem and the power spectrum around the non-linear scale. This result should hold to reasonable accuracy also for $\Lambda$CDM.


\section{Two-point correlators using the velocity}\label{s:2pt}

In this section we follow the arguments of Ref.~\cite{Pajer2013} and use the self-similarity of the equations of motion to derive the general form of the two-point correlators involving the velocity, i.e.~all possible correlators of $\delta$, $\theta$ and $\bm\omega$. While it is rather straightforward to compute the correlators of $\delta$ and $\theta$, the vorticity bears some subtleties.


\subsection{Self-similarity}\label{sec selfsim}

As pointed out in Ref.~\cite{Pajer2013}, self-similarity in an EdS universe has profound implications on the computation of correlators in the EFTofLSS. The ratio of any two physical scales must be time independent, so one can choose one representative scale and all the other will be related to it except for some constant multiplicative factor. We will use the non-linear scale $\kNL$ that separates the short- from the long-distance dynamics. Since this is the crucial ingredient for deriving the results in the subsequent sections, we will briefly review the arguments discussed in Ref.~\cite{Pajer2013}.

Setting $\Omega_m =1$, one finds that the non-linear equations of motion for the dark matter fluid in Eq.~\eqref{noEFT} possess and additional symmetry often called scaling symmetry or \textit{self-similarity}. This means that given any solution $\delta(x,\tau), \vec v (x,\tau)$, one finds a new solution by considering
\be\label{eq scaling}
 \delta'(\vec{x},\tau)\equiv \delta(\lambda_x \vec{x}, \lambda_\tau \tau) \;, \quad  {\vec{v}'}(\vec{x},\tau)\equiv \frac{\lambda_\tau}{\lambda_x} \, \vec{v}(\lambda_x \vec{x}, \lambda_\tau \tau)\,,  
\ee
with the other scalings given by $\partial_{\tau,x}\rightarrow \lambda_{\tau,x}\partial_{\tau,x}$ and $\cH\rightarrow \lambda_{\tau}\cH$. Since this is true for the non-linear theory in Eq.~\eqref{noEFT}, it must be also true for the EFT corrections that arise from integrating out short modes. Imposing self-similarity on the \eqref{ce} and \eqref{Ee}, one finds the scaling
\be
%
%
&  \left\{\dss,\dst\right\} \; \rightarrow \; \left(\frac{\lambda_\tau}{\lambda_x}\right)^2 \, \left\{\dss,\dst \right\} \;, \qquad \left\{J,\tJ\right\}(\vec{x},\tau) \; \rightarrow \;\lambda_\tau^2 \, \left\{J,\tJ\right\}(\lambda_x \vec{x}, \lambda_\tau \tau) \; .
\ee
The scaling for the LECs $c_{s,sv,bv}$ and $\chi_{1,2}$ is obviously the same as for $\dss$ and $\dst$. For the new solutions to correspond to the same cosmology as the original solution, they need to have the same initial conditions. In other words we have to check whether the initial conditions break the scaling of EdS. For self-similarity to be unbroken, the following two-point correlators must transform as
\be\label{same}
\begin{split}
\ex{\delta(\vec{x},\tau)\delta(\vec{y},\tau)}& \eq \ex{\delta'(\vec{x},\tau)\delta'(\vec{y},\tau)} \eq \ex{\delta(\lambda_{x} \vec{x}, \lambda_{\tau}\tau)\delta(\lambda_{x} \vec{y}, \lambda_{\tau}\tau)}\,,\\[1.5ex]
\ex{\delta(\vec{x},\tau)\theta(\vec{y},\tau)}& \eq \ex{\delta'(\vec{x},\tau)\theta'(\vec{y},\tau)} \eq \lambda_\tau \ex{\delta(\lambda_{x} \vec{x}, \lambda_{\tau}\tau)\theta(\lambda_{x} \vec{y}, \lambda_{\tau}\tau)}\,,\\[1.5ex]
\ex{\theta(\vec{x},\tau)\theta(\vec{y},\tau)}& \eq \ex{\theta'(\vec{x},\tau)\theta'(\vec{y},\tau)} \eq\lambda_{\tau}^{2}\ex{\theta(\lambda_{x} \vec{x}, \lambda_{\tau}\tau)\theta(\lambda_{x} \vec{y}, \lambda_{\tau}\tau)}\,,\\[1.5ex]
\ex{\omega^i(\vec{x},\tau)\omega^j(\vec{y},\tau)}&\eq\ex{{\omega'}^i(\vec{x},\tau){\omega'}^j(\vec{y},\tau)}\eq\lambda_{\tau}^{2}\ex{\omega^i(\lambda_{x} \vec{x}, \lambda_{\tau}\tau)\omega^j(\lambda_{x} \vec{y}, \lambda_{\tau}\tau)}\,.
\end{split}
\ee
To take advantage of this scaling behaviour, it is easier to construct a quantity that must be invariant under the self-similarity transformation for every correlator involving $\delta$, $\theta$ and $\bm{\omega}$. First though, we have to introduce some notation. For any scalar field $O$, such as $\delta$ or $\theta$, and any transverse vector field such as $\bm{\omega}$, we have
\be
\ex{O_1(\vec{k}) O_2(\vec{k}')} &\eq& (2\pi)^3 \delta_D^{(3)}(\vec{k}+ \vec{k}') \, P_{O_1 O_2}(k,\tau) \;, \\[1.5ex]
\ex{\omega^{i}(\vec{k}) \omega^{j}(\vec{k}')} &\eq& (2\pi)^3 \delta_D^{(3)}(\vec{k}+ \vec{k}')  \left[\delta^{ij}-\frac{k^{i}k^{j}}{k^{2}}\right] P_{\omega\omega}(k,\tau) \;,
\ee
with $k = |\vec{k}|$ and $i, j$ denoting the spacial components of a vector. Any correlation of $\bm{\omega}$ with a scalar field is zero because of statistical isotropy and $\bm{\omega}$ being transverse. It will be useful to define the quantity $\Delta_{O_1 O_2}^2$ through
\be
\Delta_{O_1 O_2}^{2}(k,\tau) \equiv \frac{k^{3}}{2\pi^{2}}P_{O_1 O_2}(k,\tau) \;, 
\ee
where $P_{O_1 O_2}$ may also be the $P_{\omega\omega}$ power spectrum. Then, from Eqs.~\eqref{eq scaling} and \eqref{same}, we find that preserving self-similarity requires
\be
\begin{split}
\Delta_{\delta\delta}^2(k,\tau) &\eq\Delta_{\delta\delta}^2(k/\lambda_x , \lambda_\tau \tau) \;, \qquad \quad
\Delta_{\delta\theta}^2(k,\tau)\eq \lambda_{\tau} \Delta_{\delta\theta}^2(k/\lambda_x , \lambda_\tau \tau) \;, \\[1.5ex]
\Delta_{\theta\theta}^2(k,\tau)&\eq \lambda_{\tau}^{2} \Delta_{\theta\theta}^2(k/\lambda_x , \lambda_\tau \tau) \;, \qquad \quad
\Delta_{\omega\omega}^2(k,\tau)\eq \lambda_{\tau}^{2} \Delta_{\omega\omega}^2(k/\lambda_x , \lambda_\tau \tau) \;.
\end{split}
\ee
In particular, this means that for any field one can define some quantity related to the power spectrum that must be invariant under self-similarity
\be
\inv_{\delta\delta}(k,\tau)&\equiv&\Delta_{\delta\delta}^2(k,\tau)\;, \phantom{\cH^{-2}}\qquad \quad  \inv_{\delta\theta}(k,\tau) \equiv -\hubble^{-1} \,  \Delta_{\delta\theta}^2(k,\tau)\;, \\[1.5ex]
\inv_{\theta\theta}(k,\tau)&\equiv&\cH^{-2} \Delta_{\theta\theta}^2(k,\tau)\,, \qquad \quad  \inv_{\omega\omega}(k,\tau) \equiv \cH^{-2} \Delta_{\omega\omega}^2(k,\tau)\;, 
\ee
i.e.~$\inv(k/\lambda_{x},\lambda_{\tau}\tau)=\inv(k,\tau)$. Since these equalities must be satisfied at all times for all scales, we can consider them on large scales, which are well described by linear physics. On these scales, inhomogeneities in the density grow linearly with the scale factor $\delta_1 \propto a $, leading to $P_{\delta\delta}\sim a^{2}$. The only way to make this compatible with self-similarity is to assume a power-law initial power spectrum $P_{\delta\delta,\textit{in}}(k)$ 
\begin{equation}\label{eq initialP}
P_{\delta\delta,\textit{in}}(k) \eq A \, k^n \;,
\end{equation}
where $A$ is an amplitude and $n$ is a spectral index. One then finds that $\inv_{\delta\delta}$ is invariant iff the scaling of $\vec{x}$ and $\tau$ is chosen such that $\lambda_x = \lambda_\tau^{4/(n+3)}$. In other words, power-law initial conditions break all but one of the infinite self-similar scalings of Eq.~\eqref{eq scaling} that leave the equations of motion invariant. In the following, we will always assume $\lambda_x = \lambda_\tau^{4/(n+3)}$ when we talk about self-similarity. Because of self-similarity the ratio of any two physical scales must be constant in time (simply because by dimensional analysis the scaling with $\lambda_{x}$ must cancel out). It is then convenient to choose a scale with respect to which we measure all others. A useful choice is the non-linear scale $k_{NL}$, which we define by $\Delta_{\delta\delta, lin}^{2}(k_{NL})=1$ and hence
\be \label{eq knl}
\kNL \eq \left( \frac{2\pi^2}{A \, a^2} \right)^{1/(n+3)} \;.
\ee
With this definition it is clear that at linear order
\be\label{eq Ilinear}
\inv_{\delta\delta,lin}(k,\tau)=\Delta_{\delta\delta,lin}^2 (k,\tau) = \left(\frac{k}{k_{NL}}\right)^{3+n}\,.
\ee
Notice that the ratio $k/k_{NL}$ is invariant under self-similarity. Hence \textit{any $\inv(k,\tau)$ can be written as function of only this ratio}. In particular, since hydrodynamics and EFT in general are constructed as an expansion in derivatives, any $\inv$ will be a polynomial in $k/k_{NL}$. For example, using the linear order relation $\theta =-\cH \delta$ one finds
\be\label{eq IlinearII}
\inv_{\delta\theta,lin}(k,\tau) \eq \inv_{\theta \theta, lin}(k,\tau) \eq \left(\frac{k}{k_{NL}}\right)^{n+3}\,.
\ee 

As opposed to the divergence $\theta$, the vorticity $\bm{\omega}$ has no direct relation to $\delta$. We can parametrize the leading term to the power spectrum of vorticity as
\begin{equation}\label{eq Pwwgeneral}
P_{\omega \omega}(k,\tau) \eq A_\omega \, a^m k^{n_w} \;, 
\end{equation}
for some amplitude $A_\omega$ and arbitrary exponents $m$ and $n_w$. Requiring the corresponding correlator $\inv_{\omega \omega}$ to be invariant under a self-similarity transformation implies a non-trivial relation between $m$, $n_w$ and the spectral index $n$ of $P_{\delta\delta, in}$ 

\begin{equation}\label{eq mnw}
m \eq 2 \, \frac{3+n_w}{3+n}-1 \;.
\end{equation}
The linear equation of motion for $\bm \omega$ predicts the decay of vorticity as $a^{-1}$ and therefore $m=-2$. But the initial power spectrum of vorticity can be arbitrary, even non-power law like, depending on what the microphysics of the early universe is that generates primordial vorticity. So we expect to have some initial violation of self-similarity (i.e.~violation of Eq.~\eqref{eq mnw}) at early time. Because of the $a^{-1}$ scaling of vorticity at linear order, at later times the primordial contribution is very suppressed and can be neglected. What determines then the leading order  vorticity are the EFT corrections, and in particular the noise term in Eq.~\eqref{eq eomlin}. As we will discuss in Sec.~\ref{s:vorticity}, this leads to a term with $n_{w}=4$, which, using Eq.~\eqref{eq mnw}, uniquely fixed the time dependence once the scalar spectral index is specified.

 
\subsection{The cancellation of divergences}\label{s:canceldiv}

In this subsection we show that all divergences in the correlators of $\delta$ and $\theta$ at one loop can be cancelled by the counterterms in the EFT terms in \eqref{eq eomlin} and \eqref{eq eomlin2}. The $\alpha$ and $\beta$ interactions in \eqref{noEFT} generate two different types of diagrams at one loop (see e.g.~\cite{Bernardeau2002}), namely $P_{13}$ and $P_{22}$. It is convenient to expand the integrands for $k\ll q$, with $k$ the external momentum and $q$ the momentum in the loop, and then perform the three dimensional integral of the leading term. We find\footnote{In the case $n=1/2$ for $P_{22}$ and $n=-1$ for $P_{13}$ these expressions are divergent. In those specific cases the angular integral must be performed \textit{before} taking the $k\ll q$ limit to obtain a sensible result.}
\be
P_{\delta_{2}\delta_{2}}&=&-\frac{a^{4}A^{2}}{2\pi^{2}} k^{4}\Lambda^{2n-1}\left[\frac{9}{98 (2n-1)}+\mathcal{O}\left(\frac{k}{\Lambda}\right)\right]\,,\\
P_{\delta_{2}\theta_{2}}&=&P_{\theta_{2}\delta_{2}}=\frac{a^{4}A^{2} \cH}{2\pi^{2}} k^{4}\Lambda^{2n-1}\left[\frac{19}{294 (2n-1)}+\mathcal{O}\left(\frac{k}{\Lambda}\right)\right]\,,\\
P_{\theta_{2}\theta_{2}}&=&-\frac{a^{4}A^{2} \cH^{2}}{2\pi^{2}} k^{4}\Lambda^{2n-1}\left[\frac{61}{490 (2n-1)}+\mathcal{O}\left(\frac{k}{\Lambda}\right)\right]\,,\\
P_{\theta_{1}\theta_{3}}&=&-P_{\delta_{1}\theta_{3}}\cH=-3 \frac{A^{2}a^{4} \cH^{2}}{4\pi^{2}}\,k^{n+2} \Lambda^{n+1} \left[ \frac{1}{5 (n+1)}+\mathcal{O}\left(\frac{k}{\Lambda}\right)\right]\,,\\
P_{\delta_{1}\delta_{3}}&=&-P_{\theta_{1}\delta_{3}} \cH^{-1}=3 \frac{A^{2}a^{4} }{4\pi^{2}}\, k^{n+2} \Lambda^{n+1}\left[\frac{ 61 }{945 (n+1)}+\mathcal{O}\left(\frac{k}{\Lambda}\right)\right]\,.
\ee
The first thing to notice is that there are five independent divergences, since some of the correlators are trivially proportional to each other. The additional correlators coming from EFT terms can be easily extracted from \eqref{eq deltacj}
\be
\ex{ \delta_c (\vec{k}) \delta_{1} (\vec k)}&=& -\cH^{-1}\ex{ \delta_c (\vec{k}) \theta_{1} (\vec k)} \supset -   \frac{  A a^{2} k^{2+n}}{\cH^{2}}\left[\frac{1}{9}\dss_{ctr}(\Lambda)-\frac{7}{18}\dst_{ctr}(\Lambda)\right] \;,\\
 \ex{\theta_c (\vec{k})\theta_{1}(\vec k)}& =&-\cH \ex{\theta_c (\vec{k})\delta_{1}(\vec k)}\supset  \frac{ A a^{2} k^{2+n}}{\cH}\left[\frac{1}{3}\dss_{ctr}(\Lambda)-\frac{1}{6}\dst_{ctr}(\Lambda)\right] \;,\\
\ex{ \delta_J (\vec{k}) \delta_{J}(\vec{k})} & \supset & \frac{1}{\cH^{4}} \left[\frac{4}{49}\ex{J_{ctr}J_{ctr}}-\frac{20}{49} \ex{J_{ctr}\tJ_{ctr}}+\frac{25}{49}\ex{\tJ_{ctr}\tJ_{ctr}}\right]\,,\\
\ex{ \delta_J (\vec{k}) \theta_{J}(\vec{k})} & \supset & -\frac{1}{\cH^{3}} \left[\frac{8}{49}\ex{J_{ctr}J_{ctr}}-\frac{26}{49} \ex{J_{ctr}\tJ_{ctr}}+\frac{15}{49}\ex{\tJ_{ctr}\tJ_{ctr}}\right]\,,\\
\ex{ \theta_J (\vec{k}) \theta_{J}(\vec{k})} & \supset & \frac{1}{\cH^{2}} \left[\frac{16}{49}\ex{J_{ctr}J_{ctr}}-\frac{-24}{49} \ex{J_{ctr}\tJ_{ctr}}+\frac{9}{49}\ex{\tJ_{ctr}\tJ_{ctr}}\right]\,.
\ee
We see that there are five counterterms, namely $\ex{J_{ctr}J_{ctr}},\ex{J_{ctr}\tJ_{ctr}},\ex{\tJ_{ctr}\tJ_{ctr}},\dss_{ctr}$ and $\dst_{ctr}$. Also, the time and $k$ dependence matches exactly with that in the loop integrals if one accounts for \eqref{vio} and the argument in Appx.~\ref{sec JJ} that leads to \eqref{ok} and hence
\be
\ex{J_{ctr}J_{ctr}}\sim\ex{J_{ctr}\tJ_{ctr}}\sim\ex{\tJ_{ctr}\tJ_{ctr}}\sim k^{4}a^{2}\,.
\ee
This means that all divergences can be cancelled exactly. Since the number of divergences is the same as the number of free parameters, unfortunately the cancellation is not a check of algebraic mistakes. Computing other correlators, such as e.g.~three-point functions, could provide such a non-trivial check. 

As anticipated, if we had not taken into account $\dst$ and $\tJ$ in the continuity equation, we would have ended up with just two parameters instead of five and the numerical factors would not allow for a cancellation of all divergences with only two LECs. It is straightforward to check that it would have been impossible to cancel all five divergences. Notice that, as long as one considers only $\delta$ as e.g.~in \cite{Carrasco2012,Hertzberg2012,Pajer2013}, there are only two divergences and so two counterterms are sufficient. At this level, i.e.~without looking at correlators involving $\theta$, the two noises $J$ and $\tJ$ and the two viscous coefficients $\dss$ and $\dst$ are degenerate with each other. Let us now move on and concentrate on the finite, physical results for these correlators.


\subsection{$P_{\delta\delta}$, $P_{\delta\theta}$, and $P_{\theta\theta}$}\label{ss:2pt}

The three two-point correlators involving the density contrast and the velocity divergence can be treated very similarly and we can follow the discussion of Ref.~\cite{Pajer2013}. Since we showed in the previous section that divergences are cancelled by the counterterms, we discuss only the finite, cutoff-independent terms in the following.
In order to get the general form of the invariants $\inv_{\delta\delta}$, $\inv_{\delta\theta}$, and $\inv_{\theta\theta}$, it is important to realize that, as consequence of self-similarity, the various contributions to these invariants have all the same form, i.e.~a numerical coefficient multiplied by some power of $k/\kNL$ (see discussion below \eqref{eq Ilinear}). It is then straightforward to find leading powers of $k/\kNL$ of the various contributions to $\inv$. As in Ref.~\cite{Pajer2013} it is most sensible to split $\inv$ into linear, loop, viscous and noise contributions

\begin{equation}\label{eq Igeneral}
\inv \eq \inv_\textit{lin} + \inv_\textit{loop} + \inv_\textit{c} + \inv_J \;,
\end{equation}
with $\inv$ standing for $\inv_{\delta\delta}$, $\inv_{\delta\theta}$, or $\inv_{\theta\theta}$. $\kNL$ has been defined in such a way that the linear part of $\inv$ is simply given by Eqs.~\eqref{eq Ilinear} and \eqref{eq IlinearII}.

 
\subsubsection*{Loop corrections}

The non-linear corrections up to $N$ loops can be calculated as in SPT. As can be seen simply through dimensional analysis, the finite parts take the form

\begin{equation}\label{eq Iloop}
\inv_\textit{loop} \eq \sum_{i=1}^N \,  \left( \frac{k}{\kNL} \right)^{(n+3)(i+1)} \left\{ \alpha_i + \tilde{\alpha}_i \ln \left( \frac{k}{\kNL} \right) \right\} \;.
\end{equation} 
In Appx.~\ref{sec formulae} the explicit form of the one-loop corrections are given for $\inv_{\delta\theta}$, $\inv_{\theta\theta}$, while the one for $\inv_{\delta\delta}$ is found in Ref.~\cite{Pajer2013}. In Tab.~\ref{tab alpha} we report the numerical values for $\alpha_1$ and $\tilde{\alpha}_1$. Let us stress again that $\alpha_1$ and $\tilde{\alpha}_1$ represent the \textit{finite} part of the integral arising in the computation of the loop corrections to the power spectra. For a more intuitive understanding of the renormalization procedure, we discussed the cancellation of divergences in the previous section using a cutoff regularization of the integrals. However, the integrals are computed most conveniently using dimensional regularization. Therefore we choose a renormalization scheme for the LECs that is similar to the $\overline{\mathrm{MS}}$ prescription commonly used in particle physics. In contrast to the MS scheme where only the divergence is absorbed in the LECs, in the $\overline{\mathrm{MS}}$ scheme we not only absorb the $1/\epsilon$ pole (where $d=3-2\epsilon$) in the LECs but also the term $\ln 4\pi - \gamma_E$ which is an artefact of dimensional regularization. This is important because whenever there is a logarithmic term\footnote{As discussed in Ref.~\cite{Pajer2013} this happens when the $k$ dependence of the loop terms coincide with the one of the EFT corrections, such as $c_{s}^{2}$, the noise or some higher derivative thereof. For example at one-loop order this happens for $(3+n)2=5+n+2m$ and $(3+n)2=7+2m$ with $m$ a positive integer, for the speed of sound and stochastic noise counterterms, respectively.}, namely $\tilde{\alpha}_1\neq 0$, $\alpha_1$ becomes degenerate with the LECs, i.e. the numerical values for $\alpha_1$ in Tab.~\ref{tab alpha} depend on the renormalization scheme (which is why this value was omitted in Ref.~\cite{Pajer2013}). Notice on the other hand that the $\tilde \alpha_i$ are universal, i.e.~any regularization and subsequent renormalization must give the same result. This can be a very useful check since e.g.~the computation in dimensional regularization bears some subtleties as we discuss in Appx.~\ref{a:dimreg}.

\begin{table}
\resizebox{0.99\textwidth}{!}{
\begin{tabular}{|c|c|c|c|c|c|c|c|c|c|c|c|}
\hline
 $n$ & $-2$ & $-3/2$ & $-1$ & $-1/2$ & 0 & 1/2 & 1 & 3/2 & 2 & 5/2 & 3 \\
 \hline \hline
 $\alpha_{1,\delta\delta}$  & 1.38 & 0.239 & 0.0489 & 0.537 & 0.336 & 0.257 & 0.00799 & $-0.0904$ & $-0.0336$ & $-0.0446$ & $-0.0213$ \\
 \hline
 $\tilde{\alpha }_{1,\delta\delta}$ & 0 & 0 & 0.194 & 0 & 0 & $-0.0918$ & $-0.0381$ & 0.00188 & 0 & $-0.0134$ & 0.0151 \\
 \hline
\end{tabular}
}

\vspace{2ex}

\resizebox{0.99\textwidth}{!}{
\begin{tabular}{|c|c|c|c|c|c|c|c|c|c|c|c|}
\hline
 $n$ & $-2$ & $-3/2$ & $-1$ & $-1/2$ & 0 & 1/2 & 1 & 3/2 & 2 & 5/2 & 3 \\
 \hline \hline
  $\alpha_{1,\delta \theta}$  & 0.655 & $-0.442$ & $-0.128$ & 0.852 & 0.495 & 0.488 & 0.0445 & $-0.231$ & $-0.0671$ & $-0.0848$ & $-0.0317$ \\
  \hline
 $\tilde{\alpha }_{1,\delta \theta}$ & 0 & 0 & 0.397 & 0 & 0 & $-0.0646$ & $-0.125$ & 0.0237 & 0 & $-0.0202$ & 0.0248 \\
 \hline
\end{tabular}
}

\vspace{2ex}

\resizebox{0.99\textwidth}{!}{
\begin{tabular}{|c|c|c|c|c|c|c|c|c|c|c|c|}
\hline
 $n$ & $-2$ & $-3/2$ & $-1$ & $-1/2$ & 0 & 1/2 & 1 & 3/2 & 2 & 5/2 & 3 \\
 \hline \hline
  $\alpha_{1,\theta \theta}$  & 0.0755 & $-1.03$ & $-0.232$ & 1.24 & 0.755 & 0.727 & $-0.0278$ & $-0.394$ & $-0.0755$ & $-0.121$ & $-0.0436$ \\
  \hline
 $\tilde{\alpha }_{1,\theta \theta}$ & 0 & 0 & 0.6 & 0 & 0 & $-0.124$ & $-0.212$ & 0.0845 & 0. & $-0.0318$ & 0.0345 \\
 \hline
\end{tabular}
}

\caption{Numerical values for the one-loop coefficients. For $\tilde{\alpha } \neq 0$ the corresponding value for $\alpha$ is degenerate with the fitting parameters $\beta$, $\gamma$ or some higher derivative terms and has been computed using an $\overline{\mathrm{MS}}$-like prescription for the cancellation of the divergences.}\label{tab alpha}
\end{table}

 
\subsubsection*{Speed of sound and viscosity corrections}

After cancelling the divergences of the loop integrals, we are left with the finite, i.e.~renormalized, contributions from the LECs. Independent of the regularization scheme that is adopted, the renormalized LECs have to satisfy self-similarity which means that the time dependence is given by

\begin{equation}\label{eq timeds}
d_\textit{ren}^{2} \; \sim \; \tilde{d}_\textit{ren}^{2} \; \propto \;  a ^{(1-n)/(3+n)}\,.
\end{equation}
Knowing this time dependence we can explicitly compute the finite contribution to $\delta$ and $\theta$ that comes from the LECs. It is given by 

\be\label{eq deltacjfin}
 \delta_c (\vec{k}, \tau)& \eq& - \int da'\; G_\delta(a, a') \left[ d_\textit{ren}^{2}(a') - \left(\frac{3}{2}+ a' \partial_{a'} \right) \tilde{d}_\textit{ren}^{2}(a') \right]\bigtriangleup \delta_{1}(a') \nonumber \\
 &=&   \frac{ \, \bigtriangleup \delta_1}{\cH^{2}}\left[ \frac{(n+3)^{2}}{10 n +46}d_{ren}^{2}- \frac{(n+3)(3n+17)}{20n+92} \tilde{d}_\textit{ren}^{2} \right] \;,\\[1.5ex]
 \theta_c (\vec{k}, \tau)& \eq& - \int da'\; G_\theta (a, a') \left[ \left(1+a'\partial_{a'}\right)\, d_{\textit{ren}}^{2}(a')- \frac{3}{2} \tilde{d}_\textit{ren}^{2}(a') \right] \bigtriangleup \delta_{1}(a') \nonumber \\
 &=& - \frac{ \bigtriangleup \delta_1}{\cH} \left[\frac{(n+7)(n+3)}{10 n +46}d_{ren}^{2} - \frac{3(n+3)^2}{20n+92} \tilde{d}_\textit{ren}^{2} \right]\;.
\ee
Using the equations above, we can easily compute the contributions to the two-point correlators that come from the LECs $d_\textit{ren}$ and $\tilde{d}_\textit{ren}$. Replacing the time dependence of Eq.~\eqref{eq timeds} with our definition of $\kNL$ of Eq.~\eqref{eq knl} it is straight forward to see that the contributions from the speed of sound and the viscosity coefficients take the following form

\begin{equation}\label{eq scalingbeta}
\inv_\textit{c} \eq \beta \, \left( \frac{k}{\kNL} \right)^{n+5} \;.
\end{equation}
The coefficient $\beta$ is now different for the three correlators and is given by\footnote{Notice that for example $\beta_{\delta\delta}$ contains contributions from both $\ex{\delta_{c}\delta}$ and $\ex{\delta\delta_{c}}$ hence the additional factor of 2 in these formulae.}

\begin{equation}\label{eq betac}
\begin{split}
\beta_{\delta\delta} & \eq    \left\{- \frac{(n+3)^{2}}{5 n +23} \frac{d_{ren,0}^{2}}{\cH_0^2} + \frac{(n+3)(3n+17)}{10n+46} \frac{\tilde{d}_{\textit{ren},0}^{2}}{\cH_0^2}  \right\} (\kNL^0)^{2}  \;, \\[1.5ex]
\beta_{\delta\theta} & \eq  \left\{ \frac{(n+3)(n+5)}{5n+23} \frac{  d_{ren,0}^{2}}{\cH_0^2} -\frac{(n+3)(3n+13)}{10n+46}  \frac{\tilde{d}_{\textit{ren},0}^{2}}{\cH_0^2}  \right\} (\kNL^0)^{2}   \;, \\[1.5ex]
\beta_{\theta\theta} & \eq  \left\{\frac{(n+7)(n+3)}{5 n +23} \frac{d_{ren,0}^{2}}{\cH_0^2} - \frac{3(n+3)^2}{10n+46} \frac{\tilde{d}_{\textit{ren},0}^{2}}{\cH_0^2}  \right\} (\kNL^0)^{2}  \;.
\end{split}
\end{equation}
where we defined the renormalized coefficients and the non-linear scale at present time $\kNL^0$ as

\begin{equation}
d_\textit{ren}^2 \eq d_{\textit{ren},0}^2\left(\frac{a}{a_0}\right)^{(1-n)/(n+3)} \eq d_{\textit{ren},0}^2 \left(\frac{\kNL^0}{\kNL}\right)^{(1-n)/2} \;,
\end{equation}
and likewise for $\tilde{d}_\textit{ren}$. Note that the LECs enter only in the combinations of ${d}_\textit{ren}$ and $\tilde{d}_\textit{ren}$. It is therefore not possible to distinguish speed of sound, shear and bulk viscosity from each other at this level. We will discuss the physical meaning of ${d}_\textit{ren}$ and $\tilde{d}_\textit{ren}$ and make contact with simulations in Sec.~\ref{s:sims}. Let us anticipate that fixing the value of $\tilde{d}_\textit{ren}$ is equivalent to choosing a definition of what we mean by $\vv$. A convenient choice is for example $\tilde{d}_\textit{ren}=0$.

 
\subsubsection*{Noise corrections}

The last contribution to $\inv$ in Eq.~\eqref{eq Igeneral} is the short scale contribution that stems from the renormalized stochastic terms $J$ and $\tilde{J}$. The contributions to $\delta$ and $\theta$ can again be computed using the time dependence of the noise terms

\begin{equation}
J_\textit{ren} \sim \tilde{J}_\textit{ren} \propto a^{(4-n)/(3+n)}\,.
\end{equation}
This dependence follows from the momentum dependence of short-scale correlators (see Appx.~\ref{sec JJ}) and self-similarity. $\delta$ and $\theta$ then get a contribution from $J_\textit{ren}$ and $\tilde{J}_\textit{ren}$ that is given by

\be \label{dJ}
\delta_J (\vec{k}, \tau)& \eq &- \int da'\; G_\delta(a, a') \left[  J_\textit{ren}(a') - \left( \frac{3}{2} + a' \partial_a' \right) \tilde{J}_\textit{ren}(a') \right] \\
&=& \frac{1}{\cH^{2}} \left[\frac{2 (n+3)^{2}}{(4-n)(3n+23)}J_{ren} - \frac{(n+3)(n+17)}{(4-n)(3n+23)} \tilde{J}_\textit{ren} \right]\,,\\
\theta_J (\vec{k}, \tau) &\eq &-\int_{0}d a'  G_{\theta}(a,a') \left[ \left(1+a'\partial_{a'}\right) J_\textit{ren}(a') + \frac{3}{2} \tilde{J}_\textit{ren} (a') \right]  \\
  &=&- \frac{ 1}{\cH} \left[\frac{14 (n+3)}{(4-n)(3n+23)}J_{ren} - \frac{3(n+3)^2}{(4-n)(3n+23)} \tilde{J}_\textit{ren} \right]\,.
\ee
Since $J_\textit{ren}$ and $\tilde{J}_\textit{ren}$ are stochastic terms, we expect $\delta_J$ and $\theta_J$ to be uncorrelated with $\delta_1$, i.e.~the only non-zero correlator arise when $\delta_J$ and $\theta_J$ are correlated among themselves. We could in principle proceed as in the case of $\delta_c$ and $\theta_c$ and compute e.g.~the contributions of $\EV{\delta_J(\vec{k})}{\delta_J(\vec{k}')}$ to $\inv_{\delta \delta}$. In the previous section, it made sense to express the coefficient $\beta$ in terms of the LECs that appear in the equations of motion. However, in the case of the noise contributions onlty the of correlators of $J_\textit{ren}$ and $\tilde{J}_\textit{ren}$ appear. Despite being unable to measure $J_\textit{ren}$ and $\tilde{J}_\textit{ren}$ separately, we may follow the argument of Ref.~\cite{Zeldovich1965,Peebles1980}. Correlators like $\EV{J_\textit{ren}}{J_\textit{ren}}$ are purely dominated by the short scale dynamics. Therefore, as discussed in detail Appx.~\ref{sec JJ}, the contribution to the power spectrum induced by these short modes scales at least as $ k^4$. This leaves us with

\begin{equation}
\inv_\textit{J} \eq \gamma \, \left( \frac{k}{\kNL} \right)^{7} \;,
\end{equation} 
with $\gamma=\gamma_{\delta\delta,\theta\delta,\theta\theta}$ being three fitting parameters, which are just three independent linear combinations of the three noise correlators $\ex{J_{\textit{ren}}J_{\textit{ren}}}$, $\ex{J_{\textit{ren}}\tJ_{\textit{ren}}}$ and $\ex{\tJ_{\textit{ren}}\tJ_{\textit{ren}}}$. In Sec.~\ref{s:sims} we will discuss how to determine these parameters and what their physical meaning is, but let us anticipate what the bottom-line is. By choosing a specific definition of $\vv$, or equivalently specifying a procedure to extract it from the simulations, one can fix two (independent linear combinations) of these three parameters to whatever value, e.g.~zero.

 
\subsubsection*{Final result}

Summarizing, after the cancellation of divergences by the counterterms, $\inv_{\delta\delta}$, $\inv_{\delta\theta}$, and $\inv_{\theta\theta}$ take all the same form which is

\begin{equation}\label{eq generalI}
\begin{split}
\inv \eq & \left( \frac{k}{\kNL} \right)^{n+3} \left\{ 1 + \left( \frac{k}{\kNL} \right)^{n+3} \left[ \alpha_1 + \tilde{\alpha}_1 \ln \frac{k}{\kNL} \right] \right\}   \\[1.5ex]
& + \beta \, \left( \frac{k}{\kNL} \right)^{n+5} + \gamma \, \left( \frac{k}{\kNL} \right)^{7} + \dots \;,
\end{split}
\end{equation}
where the ellipsis stand for two- and higher loop contributions as well as higher order terms in the derivative and field expansion. Here $\alpha_1$ and $\tilde{\alpha}_1$ are computable coefficients that depend only on $n$ while $\beta$ and $\gamma$ (different for each of the three correlators) are fitting parameters that are not determined by the effective theory. Instead, they can be extracted from fitting simulations or observations once an operational definition of $\vv$ is provided (see section Sec.~\ref{s:sims}). Note that although there are three different $\beta$ and $\gamma$, they only depend on two fitting parameter, namely~$d_\textit{ren}^2$ and $\gamma_{\delta \delta}$. We defer to the next section a physical discussion of this result.

\begin{figure}
\centering
\includegraphics[scale=0.8]{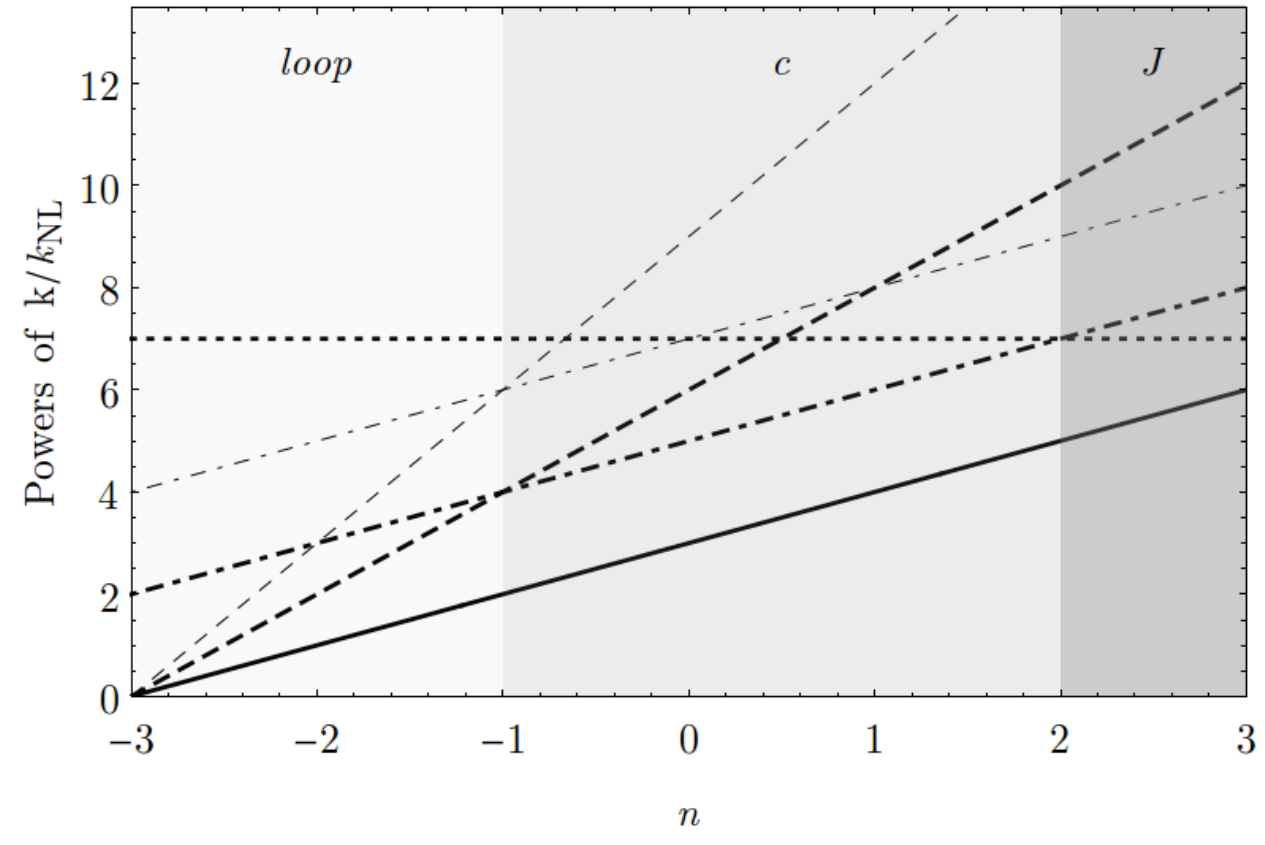}
\caption{Scaling dependence of the invariant correlator $\inv$. The solid, dashed, dot-dashed and dotted lines are the the tree-level, one-loop, LEC and noise contributions to $\inv$. In addition, we plot the scaling of the two-loop and the higher order derivative terms which are represented by the thin dashed and dot-dashed lines, respectively. The gray shaded regions delimit the regions of dominance of one of the contributions. For $n\approx-3/2$, which is relevant for our universe, we note that the LEC and one-loop contributions are much more important than the noise term.} \label{fig scalingI}
\end{figure}

One may regard the result in Eq.~\eqref{eq generalI} as the computation of the dimension of the leading irrelevant operators. The higher the exponent, the smaller the contribution for $k\ll k_{NL}$ and therefore the more irrelevant the operator. A few comments are in order. First, since the exponents of $k/\kNL$ of the various pieces in Eq.~\eqref{eq generalI} depend on $n$, the importance of different terms depends on the initial power spectrum, as has already been noticed in \cite{Pajer2013}. We compare the different scalings in Fig.~\ref{fig scalingI}, where the vertical axis is the exponent of $k/k_{NL}$ and the horizontal axis is the spectral index $n$. Second, the form of Eq.~\eqref{eq generalI} and hence Fig.~\ref{fig scalingI} is the same for all auto- and cross correlators of $\delta$ and $\theta$. This is not too surprising since $\delta$ and $\theta$ are coupled and so whatever effects contribute to one make their way into the other. Third, the interpretation of Fig.~\ref{fig scalingI} is the same as given in \cite{Pajer2013}, namely that the dissipative terms become more important than the one loop correction for $n>-2$ but are more important than two loop corrections for $n\sim -1.5$, which is relevant for our late time universe. The noise term is more subleading.

So far we neglected any contribution from the vorticity to the two-point correlators of the density contrast and the velocity divergence. In the next subsection, we will see how the vorticity leads only to terms that are suppressed by a rather high power of $k/\kNL$, i.e. $(k/\kNL)^{10+n}$ or $(k/\kNL)^{13+3n}$. This justifies neglecting it in the correlators of $\delta$ and $\theta$ as we have done in this section.

\subsection{Vorticity}\label{s:vorticity}

The case of the vorticity correlator $\inv_{\omega \omega}$ is not as closely related to $\inv_{\delta\delta}$ as the correlators with the velocity divergence. Therefore we shall investigate $\inv_{\omega \omega}$ in some detail, as well as its back reaction on $\inv_{\delta\delta}$, $\inv_{\delta\theta}$ and $\inv_{\theta\theta}$ which we neglected so far.

Assuming some initially present vorticity $\bm{\omega}_0(\vec{k})$, the linear equation of motion tells us that $\bm{\omega}$ decays as $1/a$, i.e. $P_{\omega\omega} \sim 1/a^2$. This is the reason why $P_{\omega \omega}$ is rarely discussed in the literature (see e.g.~Refs.~\cite{Pueblas2009,Scoccimarro2000} for exceptions). As can be inferred from the non-linear term in Eq.~\eqref{eq eomw} that couples $\bm{\omega}$ to $\theta$, the one-loop correction is at most constant in $a$ and only at the two-loop level we get a growing correction to the decaying linear solution. We assume that the perturbative expansion holds for the vorticity as well, i.e.~that the amplitude of the loop corrections is small enough to be significantly suppressed despite the different behaviour in time as is the case for $\delta$ and $\theta$. This is a good assumption since $\bm{\omega}$ and $\theta$ are of the same order at the shell crossing scale, but $\bm{\omega}$ decays faster than $\theta$ on larger scales. 

In the EFTofLSS the noise term $\Delta J$ in the effective stress tensor acts as constant source of vorticity, implying that $P_{\omega \omega}$ is not decaying in time even at leading order. In fact, we may assume that all primordial vorticity in the universe is already gone and that therefore the vorticity generated at short scales is dominant. The contribution to $\inv_{\omega \omega}$ from the noise term can be estimated in complete analogy to the case of $\delta$ and $\theta$ (see Appx.~\ref{sec JJ}). More specifically, since the velocity induced by short modes scales as $\vec{v} \sim \vec{k}$, $\theta$ as well as $\bm{\omega}$ scale as $\sim k^2$, leading to a $(k/\kNL)^7$ term in the invariant correlator $\inv_{\omega \omega}$.

As pointed out in Ref.~\cite{Carrasco2013a}, there is a second contribution to $\inv_{\omega \omega}$ that can be of the same order as the noise term. The $c_{sv}^2$ term in Eq.~\eqref{eq eomlin3} generates only a term that is proportional to $P_{\omega \omega}$ itself. Due to the additional derivatves this will be suppressed compared to the noise term. However, if we expand the effective stress tensor to higher orders, we will get terms that are of the form $\sim \partial^2 \delta^2$ which act as a source term in the equation of motion for $\omega$. Despite the fact that such terms are naively of higher order in the number of perturbations, they contain $\delta$ instead of $\omega$ and they can give a contribution that is of the same order as the noise term discussed previously. The correlator of two such terms gives a contribution  

\be
\ex{\omega|_{\partial^2 \delta^2} \; \omega|_{\partial^2 \delta^2}} &\sim& k_1^2 k_2^2 \, \int_{{\bf q}_1}\int_{{\bf q}_2} \EV{\delta_1({\bf q}_1) \delta_1( \vk_1 - {\bf q}_1) }{ \delta_1({\bf q}_2) \delta_1( \vk_2 - {\bf q}_2) } \non
 &\sim& \; \dirac(\vk_1 + \vk_2) \; \left[ k_1^{2n+7} +\mathrm{divergences}\right]  \;.
\ee
The divergent part can be renormalized by some counterterms (e.g.~the noise term) while finite part leads to $(k/\kNL)^{10+2n}$ in $\inv_{\omega \omega}$. This is in fact exactly of the same order as the noise contribution for $n=-3/2$. Hence, the invariant vorticity correlator takes the following form

\begin{equation}\label{eq Iww}
\inv_{\omega \omega} \eq \gamma_{\omega \omega} \left( \frac{k}{\kNL} \right)^7 + \beta_{\omega \omega} \left( \frac{k}{\kNL} \right)^{10+2n} + ... \;, 
\end{equation}
where $\gamma_{\omega \omega}$ and $\beta_{\omega \omega}$ are fitting parameters. The ellipsis stand for terms of higher order in $k/\kNL$ or terms that decay with the scale factor. This result tells us that the dominant contribution to the vorticity correlator is completely induced by the effective stress tensor, i.e.~short scale physics. Note well that also in $\Lambda$CDM the wavevector dependence will be he same, i.e.~$P_{\omega \omega} \sim k^4$ and $P_{\omega \omega} \sim k^{7+2n}$, respectively. The time dependence may change since $\Lambda$CDM enjoys only an approximate self-similarity. However, these deviations are expected to be small for redshifts $z\gtrsim 1$. 

From the equations of motion that include the vorticity (see Eq.~\eqref{eq eomw}), we notice that at second order in SPT $\bm{\omega}_{2}\sim \bm{\omega}_{1}\theta_{1} + \bm{\omega}_{1}\bm{\omega}_{1}$. When used to compute the power spectrum of $\bm{\omega}$, these perturbative non-linear effects should obey the $k^{4}$ suppression that we derived in general (i.e.~in the fully non-linear theory) in Appx.~\ref{sec JJ}. In fact one can think of these corrections as a contribution to the noise term when the short scales are integrated out, as in the discussion of subsection \ref{ss:out}. 

Let us check this $k^{4}$ scaling for the $\bm{\omega}_{1}\theta_{1}$ correction at the one-loop level. Since we will not be interested in the time dependence of the power spectrum we can safely drop all factors of $a$ and the Green's function and focus only on the $k$ structure.

\be
P_{\omega\theta\omega\theta} &\sim &  \int \frac{d^3q}{(2\pi)^3} \frac{d^3q'}{(2\pi)^3}  \; \gamma_{ij}(\vec{q},\vec{k}-\vec{q})\gamma_{il}(\vec{q}^{\, \prime},\vec{k}'-\vec{q}^{\, \prime}) \ex{\bm{\omega}^{j}(\vec{k}-\vec{q})\theta(\vec{q})\bm{\omega}(\vec{k}'-\vec{q}^{\, \prime})\theta(\vec{q}^{\, \prime})} \nonumber \\[1.5ex]
&=& \int \frac{d^3q}{(2\pi)^3} \; \gamma_{ij}(\vec{q},\vec{k}-\vec{q}) P_{\theta\theta}(\vq)P_{\omega\omega}(\vec{k}-\vec{q}\,) \left(\delta_{jl}-\frac{k_{j}k_{l}}{k^{2}}\right) \Big[ \gamma_{il}(\vec{q},\vec{k}-\vec{q}\,) \\[1.5ex]
& & +\gamma_{il}(\vec{k}-\vec{q},\vec{q}\,)  \Big]\,.\nonumber
\ee
Because of the projector onto the plane perpendicular to $\vec{k}$ we can drop the terms in $\gamma_{ij}$ that are proportional to $\vec{k}$. Then we find
\be
P_{\omega\theta\omega\theta} &\sim& \int \frac{d^3q}{(2\pi)^3} \; P_{\theta}(\vec{q})P_{\omega}(\vec{k}-\vec{q}\,) \frac{\vec{k} \cdot \vec{q}}{q^{2}} \left(\delta_{jl}-\frac{k_{j}k_{l}}{k^{2}}\right) \left[\frac{\vec{k} \cdot \vec{q}}{q^{2}}+\frac{\vec{k} \cdot (\vec{k}-\vec{q})}{|\vec{k}-\vec{q}|^{2}}  \right]\,,
\ee
which in the limit of $k\ll q$ starts at order $k^{4}$ as promised. It is exactly this scaling that allows the parameter $\gamma_{\omega \omega}$ in $\inv_{\omega \omega}$ to absorb the divergences that arise from the loop corrections to $P_{\omega\omega}$.

Unfortunately there are not many works that have investigated the vorticity using $N$-body simulations. Refs.~\cite{Pueblas2009,Scoccimarro2000} seem to support our result of having a non-decaying vorticity that is constantly generated by short modes. However, it is unclear if these simulations really reached the asymptotic value for the scaling of $P_{\omega \omega}$. It would therefore be interesting to perform a careful analysis of velocities using $N$-body simulations.

Of course, having a non-decaying vorticity implies that the correlators $P_{\delta\delta}$, $P_{\delta\theta}$ and $P_{\theta\theta}$ will get a correction from $\bm{\omega}$. Assuming the general form of $P_{\omega \omega}\propto a^m k^{n_\omega} $, we can find the scaling power of $k/\kNL$ in $\inv_{\delta\delta}$, $\inv_{\delta\theta}$ and $\inv_{\theta\theta}$ that is caused by $P_{\omega\omega}$. Such corrections can only arise at the loop level since the linear equations of motion do not yield a $\bm{\omega}$-contribution to either $\delta$ or $\theta$. In Ref.~\cite{Pueblas2009} the one-loop corrections have been computed for $P_{\delta \delta}$. Considering only the time dependence and ignoring the time integrals, the general structure of the vorticity corrections is  

\begin{equation}
\Delta P_{\delta \delta} (k,\tau) \; \sim \; \cH^{-2} P_{\delta\delta, \textit{lin} } \,  P_{\omega\omega} \; \sim \; a^{m+3} \;. 
\end{equation}  
For $\Delta P_{\delta \theta}$ and $\Delta P_{\theta \theta}$ the same relation holds up to factors of $\hubble$. Therefore, we know that there is an extra term in the invariant correlators $\inv_{\delta\delta}$, $\inv_{\delta\theta}$ and $\inv_{\theta\theta}$ that scales as 

\begin{equation}\label{eq scaleIdd}
\inv_{\delta\delta, \delta \theta,\theta\theta} \Big|_{\omega} \propto \eta(n,n_w) \left( \frac{k}{\kNL} \right)^{(3+n)(m+3)/2} \simeq \left\{ \begin{array}{l} \gamma_{\omega\omega}\eta(n,4)  \left( \frac{k}{\kNL} \right)^{10+n} \\[1.5ex] \beta_{\omega\omega}\eta(n,7+2n)  \left( \frac{k}{\kNL} \right)^{13+3n} \end{array} \right.  \;,
\end{equation}
where the coefficient $\eta(n,n_w)$ can be computed in SPT as was done for the coefficients $\alpha$ and $\tilde{\alpha}$. In the last equality we inserted the two values for $n_w$ that we obtained in Eq.~\eqref{eq Iww} and we used the relation between $n_w$, $m$ and $n$ that is induced by self-similarity (see Eq.~\eqref{eq mnw}). We can check these scalings by looking at the $k$-dependence. As can be seen from the form of the one-loop integrals in Ref.~\cite{Pueblas2009}, $\Delta P_{\delta \delta}$ scales as $\sim k^{3+n+n_w}$, confirming the scaling in Eq.~\eqref{eq scaleIdd}. Note that e.g.~for $n=-3/2$, the correction induced by the vorticity is of the same order as the $5$-loop contribution (and the two contributions in $\inv_{\omega \omega}$ are the same). We may therefore, as the results of Ref.~\cite{Pueblas2009} already suggested, safely neglect the back reaction of the vorticity on the correlators of the density and velocity divergence.


\section{Two-point correlators using the momentum}\label{s:mom}

In this section we discuss how the EFTofLSS changes if one uses momentum $\vp$ instead of velocity. First of all we need to find the correct equations of motion. We claim that they are given by 

\be
\partial_{\tau} \delta+\frac{\nabla \cdot \vp}{\bar \rho}&=&0\,, \label{ceq}\\
\partial_{\tau}\pi^{i}+4 \cH \pi^{i}+ \bar \rho \left(1+\delta\right) \partial_{i} \phi +\partial_{j} \frac{\pi^{i}\pi^{j}}{\bar \rho (1+\delta)}&=& \partial_{i}\tau^{ij}\label{eeq}\\
\bigtriangleup \phi_{l}&=& \frac{3}{2} \cH^{2}\delta\,,
\ee
with an EFT ansatz for the derivative of the stress tensor, namely an expantion in powers of derivatives and perturbations, including all operators compatible with the symmetries in Sec.~\ref{ss:sym}. At the order we are working, the first terms are 
\be\label{ex}
\partial_{i}\tau^{ij}\supset c_{1}^{2}\, \partial^{j} \delta +\frac{c_{2}^{2}}{\cH} \, \frac{\bigtriangleup \pi^j}{\bar \rho} + \frac{c_{3}^{2}}{\cH}\, \frac{\partial^j \vec{\nabla} \cdot \vp}{\bar\rho} + \Delta J_{\pi}^{j}+\dots\,,
\ee
with $J_{\pi}$ a stochastic variable and $c_{1,2,3}$ some time dependent functions. As we did for $\vv$, we give three justifications for this form of the equations of motion. First, one can derive these equations by writing all possible operators and then constrain them using the symmetries as we did in subsection \ref{ss:eom}. The additional complication now is that the symmetries cannot forbid total derivative terms in the right-hand side of the continuity equation \eqref{ceq}, such as $\bigtriangleup \delta$ and $\tilde J_{\pi}$ with $\tilde J_{\pi}$ starting at order $k^{2}$ in Fourier space. But since \eqref{ceq} is a linear equation one can reabsorb\footnote{Notice that this could not have been done for $\vv$ because \eqref{ce} is non-linear. If one tries to reabsorb the source terms in \eqref{ce} into $\vv$ for some smoothing scale $\Lambda_{1}$, these corrections will reappear for any other smoothing scale $\Lambda\neq \Lambda_{1}$. Since any physical results should be independent of $\Lambda$ and the formulae are the simplest after taking the limit $\Lambda\rightarrow \infty$, this suggests a convenient definition of $\vv$, which we will discuss in Sec.~\ref{s:sims}.} these terms in the definition of $\vp$. Then they reappear in Eq.~\eqref{eeq} but there they can in turn be reabsorbed into the already existing terms in $\partial_{i}\tau^{ij}$, and therefore the final equations takes the form \eqref{ceq} and \eqref{eeq} as anticipated. 

A second way to justify these equations of motion is to proceed in the same as in Sec.~\ref{ss:out}. We start with \eqref{ceq} and \eqref{eeq} and check whether integrating out short modes  in perturbation theory generates new couplings. It is clear that \eqref{ceq} is unaffected by integrating out short modes because it is linear (unlike \eqref{ce}). The non-linear term in \eqref{eeq} instead generates new terms, but these are exactly those we wrote down in \eqref{ex} plus terms that are of higher order in derivatives and perturbations.

Finally, a third way\footnote{Actually, this argument is not independent from the argument of integrating out short modes since the divergences come exactly from the same couplings that generate new terms once the short modes are integrated out.} to be sure that \eqref{ceq} and \eqref{eeq} are not missing any term is to check that all divergences cancel. We now show that this is the case for the power spectrum at one loop. As defined previously, we have
\be
\mu\equiv \frac{\nabla \cdot \vp}{\bar{\rho}}\,,\qquad \qquad \vn\equiv\frac{\nabla\times \vp}{\bar{\rho}}\,, 
\ee
from which we see that $\mu_{p}=-  \, \cH  p \delta_{p}$ at any order $p$ in perturbation theory. Then the power spectrum of $\mu$ from one loop corrections is simply given by

\be\label{eq Pmumuloop}
P_{\mu_{2} \mu_{2}} &\eq& 4 \hubble^2  P_{\delta_{2} \delta_{2}} \,,\qquad  \qquad P_{\mu_{1} \mu_{3}} =3 \hubble^2 P_{ \delta_{1} \delta_{3}} \,,\\
P_{\mu_{2} \delta_{2}}& \eq& -2 \hubble  P_{\delta_{2} \delta_{2}} \,,\qquad   P_{\mu_{1} \delta_{3}} =- \hubble P_{ \delta_{1} \delta_{3}}\,,\qquad   P_{\mu_{3} \delta_{1}} =-3 \hubble P_{ \delta_{1} \delta_{3}} \,.
\ee
The terms $P_{  \delta_{2} \delta_{2}}$ and $P_{ \delta_{1} \delta_{3}}$ correspond to the usual one-loop contributions to $P_{\delta \delta}$ which are defined in Appx.~\ref{sec formulae}. We know that these one-loop integrals have UV divergences for generic initial conditions, but these are cancelled by the same counterterms as for $\delta$. In fact using $\mu=-\cH a \partial_{a} \delta$ (valid at all orders) one finds the following relation between the cutoff-dependent contributions to the EFT corrections (the counterterms as opposed to the cutoff-independent renormalized couplings)
\be
\mu_{J_{ctr}}=-2 \cH  \delta_{J_{ctr}} \,,\qquad \qquad \mu_{d_{ctr}}=- 3 \cH \delta_{d_{ctr}}\,.
\ee
This is simply because the counterterms must have exactly the same time dependence as the divergent integrals they are cancelling and so the numerical coefficients are the same as in \eqref{eq Pmumuloop}. We hence see that the UV divergences in the $\mu$ correlators are cancelled with the same counterterms as those in $\delta$.

Finally, let us comment on IR divergences. Note well that for $-3< n<-1$ the IR divergences that are present in $P_{ \delta_{2} \delta_{2}}$ and $P_{  \delta_{1} \delta_{3}}$ cancel each other only in the combination that enters $P_{\delta \delta}$, i.e.~$P_{ \delta_{2} \delta_{2}} + P_{ \delta_{1} \delta_{3}}+ P_{ \delta_{3} \delta_{1}}$. The one-loop $\mu$ power spectrum however is given by
\be
P_{\mu\mu}|_{1-loop}=\cH^{2} \left[4P_{\delta_{2}\delta_{2}}+3P_{\delta_{1}\delta_{3}} +3P_{\delta_{3}\delta_{1}} \right]\,.
\ee
The IR divergences do not cancel due to the different numerical factors and $P_{\mu \mu}$ remains with a residual IR divergence. Although this might be surprising at first sight, it is actually what we should expect. As shown in Ref.~\cite{Jain1996}, the invariance under Galileo transformations of $\delta$ and $\theta$ ensures IR safety for the correlators of $\delta$ and $\theta$ for $n>-3$. However, $\mu$ is not invariant under a Galilean transformation since under a constant boost $\vec{u}$ one finds that $\vec{v}$ transforms as $\vec{v} \rightarrow \vec{v} - \vec{u}$ and therefore $\mu$ transforms as $\mu \rightarrow \mu - \vec{u}\cdot \vec{\nabla}\rho
$ while $\theta$ remains invariant. This means that $P_{\mu\mu}$ has IR divergences for $n<-1$ (see also Ref.~\cite{Scoccimarro1996} for a nice discussion of IR divergences in SPT)\footnote{The issue of IR divergences has received considerable attention recently \cite{Kehagias2013,Peloso2013,Blas2013}. Already in the original papers on Reg PT \cite{Crocce2006,Crocce2006a} it was realized that the common resummation schemes are affected by non-physical IR divergences that appear at higher orders. Our results, however, are not affected by this discussion since we do not attempt to include partial higher-order contributions and all IR divergences cancel correctly when considering $\delta$, $\theta$ or $\bm{\omega}$ for $n>-3$.}.


\section{Connection with simulations}\label{s:sims}

Having seen that we can formulate the EFTofLSS with either $\vp$ or $\vv$ and that in the case of $\vp$ only $d^2$ and $J$ are needed to cancel all divergences (neglecting vorticity), we can already infer that the velocity counterterms are only needed to cancel the divergences in the velocity correlators but their finite part does not have any physical effect. As we shall see, the finite part of $\tilde{d}$ and $\tilde{J}$ is part of the actual definition of $\vv$. In this section, we clarify these points, namely the physical meaning of the EFT parameters $\tilde d$ and $\tJ$, and we present a cursory discussion of how to measure velocity correlators in simulations.

We have seen that the EFTofLSS can be formulated in terms of $\vp$ or $\vv$. The two approaches are equivalent being simply related according to \eqref{eq defpi}. When using momentum and neglecting the vorticity, the self and cross power spectra of $\mu$ and $\delta$ are related to the power spectrum of $\delta$ (since $\mu=-\partial_{\tau}\delta$ at all orders) and hence contain only two fitting parameters\footnote{Of course, as we have discussed, whether one can include for consistency zero, one or two parameters depends on $n$ and the number of loop corrections under consideration as depicted in Fig.~\ref{fig scalingI}.} at the order that we have considered here, namely $\beta_{\delta\delta}$ (or equivalently the linear combination of $d^{2}$ and $\tilde d^{2}$ given in \eqref{eq betac}) and $\gamma_{\delta\delta}$ (equivalently a specific linear combination of $\EV{J_\textit{ren}}{J_\textit{ren}}$, $\EV{J_\textit{ren}}{\tJ_\textit{ren}}$ and $\EV{\tJ_\textit{ren}}{\tJ_\textit{ren}}$ that is easily deducible from \eqref{dJ}). Since this is equivalent to the EFTofLSS using the velocity, one should wonder about the meaning of the five parameters we have encountered in Sec.~\ref{ss:2pt}, namely $d^{2}_{\textit{ren}}$, $\tilde d^{2}_{\textit{ren}}$, $\EV{J_\textit{ren}}{J_\textit{ren}}$, $\EV{J_\textit{ren}}{\tJ_\textit{ren}}$ and $\EV{\tJ_\textit{ren}}{\tJ_\textit{ren}}$. The resolution of this mismatch is that fixing the value of three of these five parameters, namely $\tilde d^{2}_{\textit{ren}}$, $\EV{J_\textit{ren}}{\tJ_\textit{ren}}$ and $\EV{\tJ_\textit{ren}}{\tJ_\textit{ren}}$, is tantamount to choosing a definition of $\vv$. Since we can work with whatever variables we please, these three numbers do not have any physical meaning, rather they give us the freedom to use different ``velocities''. This is best understood looking again at the relation \eqref{eq defpi}, which, up to higher order terms, can be inverted to give (again we are neglecting $\vw$)
\be 
\vv  = \frac{\vp}{\rho}- \frac{\chi_1}{\hubble} \, \vec{\nabla} \cdot \delta + \frac{\chi_2}{\hubble^2} \, \vec{\nabla} \cdot \theta - \frac{\tilde{\vec{J}} }{\hubble} + ... \;.
\ee
As we have discussed in Sec.~\ref{ss:2pt}, $\vv$ is a renormalized operator if one uses the counterterms in the right hand side to cancel the divergences appearing in the loop corrections generated by the non-linear composite operator $\vp/\rho$. To explain more in detail and to make contact with observations, we make the cutoff scale explicit and combine $\chi_{1,2}$ into $\tilde d^{2}$, which is valid at leading order
\be\label{vvdef}
\vv  = \frac{[\vp]_{\Lambda}}{[\rho]_{\Lambda}}- \frac{\left[ \tilde d^{2}_{\textit{ctr}}(\Lambda)+ \tilde d^{2}_{\textit{ren}}\right]}{\hubble} \, \vec{\nabla} \cdot [\delta]_{\Lambda}  - \frac{\tilde{\vec{J}}_{\textit{ctr}}(\Lambda)+\tilde{\vec{J}}_{\textit{ren}} }{\hubble} + ... \;.
\ee
Here $\vp$ and $\delta$ can be easily extracted from a simulation and $[.]_{\Lambda}$ denotes the smoothing on a spatial region of size $\Lambda^{-1}$. It is clear that we can take arbitrary values for $\dst_{\textit{ren}}$ and $\tilde{\vec{J}}_{\textit{ren}}$ without spoiling the cancellation of divergences, so we can try to make the most convenient choice. Imagine to compute $[\vp]_{\Lambda}/[\rho]_{\Lambda}$ from a simulation at some long wavelength $k\ll k_{NL}$ for some $\Lambda$. The dependence of this quantity on $\Lambda$ for $\Lambda\rightarrow \infty$ must be cancelled by $ \tilde d^{2}_{\textit{ctr}}(\Lambda)$ and $\tilde{\vec{J}}_{\textit{ctr}}(\Lambda)$. According to our results in Sec.~\ref{s:canceldiv}, these scale as $\tilde d^{2}_{\textit{ctr}}\sim \Lambda^{n+1}$ and $\tilde{\vec{J}}_{\textit{ctr}}\sim \Lambda^{(2n-1)/2}$. For universes of phenomenological relevance, $n\lesssim -1$, so we expect that $[\vp]_{\Lambda}/[\rho]_{\Lambda}$ evaluated at large scales will converge to some constant configuration as $\Lambda$ is increased. It is then convenient to use a velocity $\vv$ defined by $\dst(\Lambda\rightarrow \infty)=0$ and $\tilde{\vec{J}}(\Lambda\rightarrow \infty)=0$. Since all correlators of $\vv$ in Eq.~\eqref{vvdef} are independent of $\Lambda$ (i.e.~we have renormalized the theory), this procedure gives an explicit way to extract $\vv$ from a simulation. Of course in simulations one cannot really reach $\Lambda\rightarrow \infty$ due to the finite resolution, but one can easily test for convergence of the above procedure comparing simulations with increasing resolution.

Although the numerical implementation might be more involved, in principle there is no obstruction to generalize this procedure to scaling cosmologies with $-1<n<1/2$, i.e.~as long as the noise counterterm vanishes in the $\Lambda\rightarrow \infty$ limit. In this regime, $[\vp]_{\Lambda}/[\rho]_{\Lambda}$ at large scales diverges as $\Lambda\rightarrow \infty$, but at leading order in $k/k_{NL}$ this divergence should just be proportional to $\nabla \delta$ and hence can be appropriately subtracted. The convenient definition of $\vv$ is again $\dst_{\textit{ren}}=0$ and $\tilde{\vec{J}}_{\textit{ren}}=0$.

\section{Conclusions}

The EFTofLSS \cite{Baumann2012, Carrasco2012, Hertzberg2012,Pajer2013} has emerged as a very promising framework to study analytically the large-scale inhomogeneities of matter. The key feature is that the EFT approach allows for a consistent treatment of mildly non-linear perturbations by taking into account the effects that the short-scale dynamics has on larger scales. It has been pointed out in the literature that the procedure of integrating out the short-scale dynamics solves a series of problems inherent to the SPT approach: the (unsmoothed) density contrast is not a suitable expansion parameter as it can diverge due e.g.~shell-crossing, large scale matter inhomogeneities do not evolve as perfect pressureless fluid, instead feature all dissipative coefficients compatible with the symmetry of the problem and these are needed to renormalize the divergences that arise when non-linear corrections are computed.

We elucidate the difference between mass- and volume-weighted velocity and notice that it is the former that is used in the EFTofLSS. For the mass-weighted velocity the influence of short-scale modes on long scales is such that the power spectrum of short modes scales as $P \sim k^4$ for $k\rightarrow 0$. This is analogous to the well-known argument concerning the density contrast. This result is interesting for the vorticity of the velocity. In linear theory vorticity decays as $\sim 1/a$, but the noise term in the effective stress tensor arising from integrating out short scales sources the vorticity directly. We deduced the scaling property of this residual vorticity from the general scaling of the short-scale modes of the velocity and verified that the back reaction of the vorticity on the density and velocity divergence is strongly suppressed. The second leading contribution to the vorticity comes from higher order terms in the effective stress tensor. It is therefore remarkable that the scaling of the dominant contributions to the vorticity correlator are determined by the symmetries and the power spectrum around the non-linear scale.

Furthermore, we considered all two-point correlators, extending the analysis of Ref.~\cite{Pajer2013} to include the velocity divergence $\theta$. We showed that additional counterterms are needed in order to cancel all divergences that arise when considering the correlators of $\theta$. These new terms are not prohibited by any symmetry and they are generated already by integrating out short scales in perturbation theory.  In total there are five independent one-loop UV-divergences (three in the 22 and two in the 13 correlators of $\delta$ and $\theta$) and five independent counterterms, namely $d^{2}$, $\dst$ and the three 2-point correlators of $J$ and $\tJ$. The finite part of the new counterterms $\dst$ and $\tJ$ can be chosen arbitrarily (while the cutoff dependent part is fixed by the cancellation of divergences within a certain renormalization procedure) because they are part of the definition of the velocity itself. We have given an explicit example of this by discussing how to extract the velocity from simulations. The renormalized two-point correlators of $\theta$ have the same structure as the density power spectrum discussed in \cite{Pajer2013}, namely \eqref{eq generalI}. The various coefficient are different though. $\alpha$'s and $\tilde \alpha$'s (one for function of $n$ each of the three correlators) are given in table \ref{tab alpha}. The three $\beta$'s are specified in terms of just two parameter, namely $d^{2}$ and $\dst$. Together with the three $\gamma$, which should be included or not depending on the value of $n$ and the number of loops included, they give five total parameters. Three of these can be fixed by the definition of velocity as discussed in the previous section.

Finally, we commented on using the momentum $\vp$ of the effective fluid instead of the velocity field. We showed that the equations of motion are simpler when written in terms of $\delta $ and $\vp$ rather than $\vv$. The continuity equation in terms of $\vp$ takes the very simple for \eqref{ceq}. We show that all one-loop divergences in the power spectra of $\delta$ and $\vp$ can be canceled by the counterterms in the Euler equation, confirming the fact that there are only two physical fitting parameter in the theory. There are two main drawbacks of using $\vp$: the perturbation theory is more cumbersome and momentum correlators are not protected by Galilean invariance and have therefore IR divergences for spectral tilts $n\leq -1$ (rather than for $n\leq -3$ as for $\vv$). We closed with some comments on how to relate the velocity and momentum correlators to simulations. The extension of our results to a $\Lambda$CDM universe is left to future investigations.

The problems discussed in this paper leave ample room for further development. It would be interesting to investigate using $N$-body simulations the $k^{4}$ scaling of the vorticity power spectrum \eqref{res1} as well as the different scaling of mass- and volume-weighted velocities. Another avenue is to extend our analysis to the bispectrum, for which some consistency checks of the renormalization procedure should be available.


\section*{Acknowledgements}

We would like to thank Guido D'Amico, Roman Scoccimarro, David Spergel and Matias Zaldarriaga for very useful discussions and comments. L.M.~is supported by a grant form the Swiss National Science Foundation and E.P.~ is supported in part by the Department of Energy grant DE-FG02-91ER-40671.

\appendix


\section{Equations for the density and the velocity} \label{sec fullEOM}

In this appendix we write down the fully non-linear equations of motion including the vorticity and neglecting the EFT corrections that are discussed in the main text.

We define the divergence and curl of the velocity (we raise and lower all indices with $\delta_{ij}$)
\be
\theta\equiv \partial_{i} v^{i}\,,\quad \omega^{i}\equiv \epsilon^{ijk}\partial_{j} v^{k}\,,
\ee
from which it follows that $\partial_{i} w^{i}=0$. In Fourier space one finds
\be
v^{i}(\vec k)=i \frac{\epsilon^{ijj'}k_{j} w_{j'}(\vec k)}{k^{2}} -i\frac{k^{i}}{k^{2}}\theta(\vec k)\,.
\ee
The equations of motion in terms of velocity gradient and curl in momentum space are given by
\be
\partial_{\tau} \delta+ \theta &=& \int_\vq \Big[-\alpha(\vec q,\vec k-\vec q) \theta(\vec q) + \va(\vec q,\vec k-\vec q) \cdot \bm{\omega}(\vec q) \Big] \delta(\vec k-\vec q)\,, \label{eq eomd}\\[1.5ex]
\partial_{\tau} \theta+ \cH \theta+\frac{3}{2} \cH^{2}\Omega_{m} \delta&=& -\int_\vq \Big[\beta(\vec q,\vec k-\vec q) \theta(\vec q)\theta(\vec k-\vec q)+ \vb(\vec q,\vec k-\vec q) \cdot\bm{\omega}(\vec k-\vec q) \theta(\vec q) \non&&\quad  + \bm{\omega} (\vec q)\mb (\vec q,\vec k-\vec q) \bm{\omega}(\vec k-\vec q)\Big]\\[1.5ex]
\partial_{\tau} \bm{\omega} +\cH \bm{\omega} &=&\int_\vq  \Big[ \mg_{ij}(\vec q,\vec k-\vec q) \omega^{j}(\vec k-\vec q) \theta(\vec q)+\va(\vec q,\vec k-\vec q) \bm{\omega}(\vec q)\cdot  \bm{\omega}(\vec k-\vec q)\non
&&\quad +\vg_{ijl}(\vec q,\vec k-\vec q) \omega^{l}(\vec k-\vec q) \omega^{j}(\vec q) \Big] \,, \label{eq eomw}
\ee
where
\be
\alpha (\vec q_{1},\vec q_{2})&\equiv&\frac{(\vec q_{1}+\vec q_{2})\cdot \vec q_{1}}{q_{1}^{2}} \,, \qquad \va  (\vec q_{1},\vec q_{2})\equiv \frac{\vec q_{2} \times \vec q_{1}}{q_{1}^{2}}\,,\\
\beta(\vec q_{1},\vec q_{2})&\equiv &\frac{ \left(\vec q_{1} +\vec q_{2}\right)^{2}\vec q_{1} \cdot \vec q_{2} }{2 q_{1}^{2}q_{2}^{2}}\,,\qquad \vb(\vec q_{1},\vec q_{2})\equiv\frac{\vec q_{1}\times \vec q_{2}}{q_{2}^{2}}\,,\\
\mb_{ij}(\vec q_{1},\vec q_{2})&\equiv &\frac{ \left(\vec q_{1}\times \vec q_{2}\right)_{i}\left(\vec q_{1}\times \vec q_{2}\right)_{j}}{q_{1}^{2}q_{2}^{2}}\,,\\ \mg_{ij}(\vec q_{1},\vec q_{2})&\equiv& \frac{\vec q_{1,i} \left(\vec q_{1}+\vec q_{2}\right)_{j}}{q_{1}^{2}}-\frac{\left(\vec q_{1}+\vec q_{2}\right)\cdot q_{1}}{q_{1}^{2} } \delta_{ij} \\
\vg_{ijl}(\vec q_{1},\vec q_{2})&\equiv &\frac{\epsilon^{il'l} (\vec q_{1}+\vec q_{2})_{l'} q_{1}^{j}}{q_{1}^{2}}\,,
\ee
and we used the vector identity
\be
\vec \partial\times \left(v^{k}\partial_{k} \vec v\right)=- \vec \partial \times \left[\vec v \times \left(\vec \partial\times \vec v\right)\right]\,.
\ee
We rewrite schematically these equations as
\be
\partial_{\tau} \delta+ \theta &=& -\alpha \, \theta \, \delta + \va \cdot \bm{\omega} \, \delta\,,\\
\partial_{\tau} \theta+ \cH \theta+\frac{3}{2} \cH^{2}\Omega_{m} \delta&=& - \beta \, \theta \, \theta + \vb \cdot \bm{\omega} \, \theta +\mb_{ij} \omega_{i}   \omega_{j} \\
\partial_{\tau} \bm{\omega}  +\cH \bm{\omega}&=& \mg_{ij}  w^{j} \theta +\va  \bm{\omega} \cdot  \bm{\omega} +\vg_{ijl}  \omega^{l}  \omega^{j}\,,\label{weq}
\ee
This seems to disagrees with Eq.~(53) of \cite{Pueblas2009}.

 
\section{The effect of short modes on large scales} \label{sec JJ}

In this appendix we show that the contribution of the perturbations on short scales to the density, divergence and vorticity power spectra is suppressed at large scales by $(k/k_{NL})^{4}$. For density perturbations, this fact was already well-known and discussed for example (in a slightly different form) in \cite{Zeldovich1965,Peebles1980}. To the best of our knowledge a similar result for the velocity had not yet been derived in the literature. We discuss density and velocity in turn; a perturbative verification of the constraint has been given in Sec.~\ref{s:vorticity}.

 
\subsection{Smooth picture} 

We can use either a smooth description of the density fields or the particle picture, equivalently. We show both derivations, starting with the smooth description. Consider the Fourier transform of the density
\be
\delta (\vk) =\intd{\vx} \; e^{-i\vk \cdot \vx} \delta (\vx)\,,
\ee
where we have chosen $\delta(\vx)\equiv \rho(\vx)/\bar\rho-1$. Now we want to see how much the large-scale power spectrum can change if short scales evolve in some complicated way. The only thing we will assume about the laws governing the short scales is that they conserve mass and momentum. To capture this situation in real space we rearrange some initial distribution of density $\delta^{(1)}$ into a new density field $\delta^{(2)}$ that differs from $\delta^{(1)}$ only inside a small region $B$ of size $\DB$ centred around $\vxB$, while being identical outside. Since this region is supposed to capture the effects of the non-linear dynamics, we should consider $\DB\sim k_{NL}^{-1}$. Mass and momentum conservation then imply
\be
\int d^{3}x \delta^{(1)}(\vx)&=&\int d^{3}x \; \delta^{(2)}(\vx)\,,\label{com1}\\
\intd{x} \; \vx\, \delta^{(1)}(\vx)&=&\intd{x}\; \vx\, \delta^{(2)}(\vx)\,.\label{como1}
\ee
To see that the second equation is true, let us split the integral between inside and outside of the small region $B$. Outside $\delta^{(1)}(\vx)=\delta^{(2)}(\vx)$, hence we have to worry only about the inside contribution. Remember that $\int d^3x\, \vx \rho(\vx)$ is the position of the center of mass of the distribution whose dynamics is only affected by forces external to the distribution itself if momentum is conserved. In other words, no matter which momentum-conserving dynamics dictates the evolution of the short scales, the motion of the center of mass has to be the same leading to \eqref{como1}.

We wish now to compute $\delta^{(1)}-\delta^{(2)}$ at large scale in Fourier space
\be
\delta^{(1)} (\vk)-\delta^{(2)} (\vk)&=&\int  d^{3}x \, e^{-i \vk \cdot \vx}  \left[\delta^{(1)} (\vx)-\delta^{(2)}(\vx)\right] \label{ksq} \\
&=&\int_{0}^{\DB}  d^{3}y \, e^{-i \vk \cdot (\vy-\vxB)}  \left[\delta^{(1)} (\vy-\vxB)-\delta^{(2)}(\vy-\vxB)\right] \nonumber\\
&\simeq&e^{-i \vk \cdot \vxB}\int_{0}^{\DB} d^{3}y\, \left[1-i \vk \cdot \vy+|\vk \cdot \vy|^{2}+\dots \right] \left[\delta^{(1)} (\vy-\vxB) \right. \nonumber \\
& \phantom{\simeq} & \left. -\delta^{(2)}(\vy-\vxB)\right]\nonumber\\
&=&e^{-i \vk \cdot \vxB}\int_{0}^{\DB} d^{3}y \, \left[|\vk \cdot \vy|^{2}+\dots \right] \left[\delta^{(1)} (\vy-\vxB)-\delta^{(2)}(\vy-\vxB)\right]\non
&\sim& e^{-i \vk \cdot \vxB} \mathcal{O} \left(k^{2}\right)\,,
\ee
where in third line we expanded assuming large scales, namely $k\DB\ll 1$ and in the fourth line we used \eqref{com1} and \eqref{como1}. Because of statistical homogeneity or equivalently conservation of momentum, the power spectrum is proportional to $\delta_{D}^{3}(\vk+\vk')$ and hence the phase $e^{-i \vk \cdot \vxB}$ cancels out and can be dropped. Then \eqref{ksq} shows that the contributions of short scales to $\delta(\vk)$ for $k\rightarrow 0$ start quadratic order in $k$ and therefore the power spectrum must be suppressed by at least a factor $(k/k_{NL})^{4}$ in the same limit.

 
\subsection{Particle picture}

Let us now provide an equivalent derivation using the particle picture. Consider particles distributed at positions $\vx_{i}$ with velocities $\vv_{i}$ and masses $m_{i}$. The (un-smoothed) density and momentum fields are given by
\be
\rho(\vx)=\sum_{i} m_{i} \delta_{D}\left(\vx-\vx_{i}\right)\,,\quad \vp(\vx)=\sum_{i}m_{i} \vv_{i} \delta_{D}(\vx-\vx_{i})\,.
\ee
Smoothing with some filter function $\WL$, e.g.~a normalized Gaussian of variance $\Lambda^{-2}$, we find (reintroducing the labels $l$ to refer to smoothed fields for clarity)
\be
\rho_{l}(\vx)=\sum_{i} m_{i} \WL\left(\vx-\vx_{i}\right)\,,\quad \vp_{l}(\vx)=\sum_{i}m_{i} \vv_{i} \WL(\vx-\vx_{i})\,.
\ee
The perturbations are defined as always as $\delta\equiv \rho/\bar\rho-1$ and their Fourier transform for $k\neq 0$ is
\be
\delta(\vk)=\sum_{i}m_{i} e^{i\vk\cdot \vx_{i}}  \WL(\vk)\,,
\ee
where with an abuse of notation we indicate by $\WL(\vk)$ the Fourier transform of $\WL(\vx)$. We want to see again how much the power spectrum can be changed if the short-scale dynamics reshuffle particles in a complicated way. In order to do this, consider a small region $B$ as above, of size $\DB\sim k_{NL}^{-1}$ and centered around $\vxB$. As we saw previously the center position of the box appears only in an overall phase $e^{i\vk\cdot \vxB}$ that cancels out in the power spectrum because of momentum conservation. To simplify the computation, in the following, without loss of generality, we will set $\vxB=0$. Notice that this implies $|\vx_{i}|\leq\DB\ll k^{-1}$ for $i\in B$. Imagine to start from some configuration $\{\vxu_{i},\vvu_{i}\}$ and change the position and velocity of the particles inside $B$ while conserving mass and momentum, to end up with the configuration $\{\vxt_{i},\vvt_{i}\}$. Since the particles outside of the small region $B$ are unchanged, we have $\{\vxu_{i},\vvu_{i}\}=\{\vxt_{i},\vvt_{i}\}$ for $i\notin B$. Because of the conservation of mass
\be\label{com}
\sum_{i \in B}m_{i}^{(1)}=\sum_{i\in B}m_{i}^{(2)}\,,
\ee
while conservation of momentum ensures that
\be\label{como}
\sum_{i \in B}m_{i}^{(1)} \vxu_{i}&=&\sum_{i\in B}m_{i}^{(2)} \vxt_{i}\,,\\
\sum_{i \in B}m_{i}^{(1)} \vvu_{i}&=&\sum_{i\in B}m_{i}^{(2)} \vvt_{i}\,.
\ee
These last two equations (the second being the time derivative of the first) say that the position and velocity of the center of mass cannot be changed by internal, momentum-conserving forces.

 
\subsubsection{Density}

We can now proceed analogously to the smooth case and compute
\be
\delta_{l}^{(1)} (\vk)-\delta_{l}^{(2)} (\vk)&=&\sum_{i} m_{i} e^{-i\vk\cdot \vxu_{i}}\WL (\vk)-\sum_{i} m_{i} e^{-i\vk\cdot \vxt_{i}}\WL (\vk)\non
&=&\sum_{i \in B} m_{i} \WL (\vk) \left[ \left(1-i\vk\cdot \vxu_{i}+\dots \right)-  \left(1-i\vk\cdot \vxt_{i}+\dots \right)\right]\non
&\simeq& \mathcal{O}(k^{2})\,,\label{ksqsup}
\ee
where in the last line we used conservation of mass \eqref{com} and momentum \eqref{como} and the fact that $\WL(\vk)\rightarrow {\rm constant}$ for $\vk\rightarrow 0$. The $k^{2}$ suppression in \eqref{ksqsup} implies that the short-scale corrections to the power spectrum at large scales start at order $k^{4}$, in agreement with the previous derivation and section 58 of \cite{Peebles1980}. It should be noticed that this result remains unchanged in the limit in which the smoothing is removed, since for $\Lambda\rightarrow \infty$ one finds $\WL(k)\rightarrow $ constant (i.e.~just the Fourier transform of a delta function up to a phase).

 
\subsubsection{Velocity}

We will now prove an analogous result for the velocity. Things now are more subtle because a different result is obtained depending on whether the smoothing is present, $\Lambda \DB\ll 1$ or is removed, $\Lambda \DB \rightarrow \infty$. When the smoothing is over scales larger than the non-linear scale, $\Lambda\ll \DB^{-1} \sim k_{NL}$ we are properly using the mass-weighted velocity. For the mass-weighted velocity the short-scale effects to the large-scale velocity field are suppressed by at least $k^{2}$ (therefore $k^{4}$ for the power spectrum of the divergence and the vorticity) in the $k\rightarrow 0$ limit, assuming conservation of mass and momentum. 
On the other hand, as we remove the smoothing by taking the $\Lambda \DB \sim \Lambda/k_{NL}\rightarrow \infty$ limit, the mass-weighted velocity becomes the same as the volume weighted velocity, as can be easily see by substituting $W_{mass, vol}$ with delta functions in the definitions \eqref{mwv} and \eqref{vwv} and comparing them. We will show that for the volume-weighted velocity the short scale effect can lead to corrections that are unsuppressed, i.e.~$k^{0}$, therefore leading to a suppression of just $k^{2}$ for the power spectrum of the volume-weighted divergence and the vorticity. We discuss these results and their implications in subsection \ref{ss:Pa}.

As a warm-up, let us start considering the momentum. The effect of short scales is isolated as before by comparing two different configurations 
\be
\vp^{(1)}(\vk)-\vp^{(2)}(\vk)&=&\sum_{i} m_{i} \vvu_{i} e^{-i\vk\cdot \vxu_{i}}\WL (\vk)-\sum_{i} m_{i}\vvt_{i} e^{-i\vk\cdot \vxt_{i}}\WL (\vk)\non
&=&\sum_{i \in B} m_{i} \WL (\vk) \left[ \vvu_{i} \left(1-i\vk\cdot \vxu_{i}+\dots \right)- \vvt_{i} \left(1-i\vk\cdot \vxt_{i}+\dots \right)\right]\non
&\simeq& \mathcal{O}(k)\,,\label{ksqsup2}
\ee
where in the last line we used the conservation of momentum \eqref{como} and $\WL(\vk \rightarrow 0)\simeq$ constant. This result implies that the divergence and vorticity of the momentum each start at order $k^{2}$ and hence the short-scale corrections to their large-scale power spectrum must start at order $k^{4}$. What about the velocity? For that we have to work a bit harder. We start from the definition
\be
\vv_{l}=\frac{\vp_{l}}{\rho_{l}}=\frac{\sum_{i}m_{i}\vv_{i}\WL(\vx-\vx_{i})}{\sum_{i}m_{i}\WL(\vx-\vx_{i})}\,.
\ee
In order to make some progress we expand in powers of the density perturbations
\be
\bar\rho \, \vv_{l}=\frac{\vp_{l}}{1+\delta_{l}}\simeq \vp_{l}\left(1-\delta_{l}+\dots\right)\,.
\ee
As we have seen above, $\vp$ from short scales starts at order $k$ because of conservation of momentum, so we concentrate on the second term $\vp_{l}\delta_{l}$. For $k\neq 0$ its Fourier transform is given by
\be
\bar\rho^{\,-2}\vv_{l}(\vk)&\supset&\sum_{ij}m_{i}\vv_{i}m_{j}\int d^{3}x \, e^{-i\vk\cdot \vx} \WL(\vx-\vx_{i})\WL(\vx-\vx_{j})\,.
\ee
For concreteness we assume that $\WL(\vx)\propto \exp \left(-\Lambda^{2}x^{2}/2\right)$ is a normalized Gaussian, then
\be
\bar\rho^{-2}\vv_{l}(\vk)&\supset&\sum_{ij}m_{i}\vv_{i}m_{j} e^{-i\vk \cdot \vx_{i}} e^{-\Lambda^{2}|\vx_{i}-\vx_{j}|/2}\,,
\ee
up to some overall $k$-independent normalization. To isolate the effects coming from short scales we use the same trick as before and compare two different configurations
\be
\bar\rho^{-2} \left[\vvu_{l}(\vk)-\vvt_{l}(\vk)\right]&\simeq&\sum_{ij}m_{i}m_{j}\left[\vvu_{i} e^{-\Lambda^{2}|\vxu_{i}-\vxu_{j}|/2}-\vvt_{i} e^{-\Lambda^{2}|\vxt_{i}-\vxt_{j}|^{2}/2}\right]+\mathcal{O}(k)\,.\nonumber
\ee

We can divide the double sum in four terms depending on whether $i$ and $j$ are inside or outside $B$. When $i,j\notin B$ then the two configuration are identical and hence cancel out. When $i$ or $j$ or both are inside $B$ we want to consider the useful limit $\Lambda\rightarrow \infty$. Modulo proper renormalization, this is the limit we used in the rest of the paper. In this limit the Gaussians become delta functions that impose $i=j$. This eliminates one of the sums leaving
\be
\bar\rho^{-2} \left[\vvu_{l}(\vk)-\vvt_{l}(\vk)\right]&\simeq&\sum_{i}m_{i}m_{j}\left[\vvu_{i} -\vvt_{i} \right]+\mathcal{O}(k)\,.\nonumber
\ee
In the case that all masses are equal $m_{i}=m$ for every $i$, this cancels exactly by momentum conservation. Equal masses are considered in the large majority of simulations. We leave to the future a detailed investigation of what happens for un-equal masses. 



\section{Useful formulae} 

Let us collect some useful analytic expression that have been discussed in this paper.

\subsection{The smoothing of non-linear functions}\label{a:smooth}

Following \cite{Baumann2012}, we have introduced the notation $\smo{f}$ for the smoothing of an arbitrary function $f$. Then let us then define its long modes (equivalently smoothing) as $f_{l}\equiv \smo{f}$ and its short modes as $f_{s}\equiv f-f_{l}$. It is then straightforward to check the result derived in \cite{Baumann2012} for the smoothing of a product of two functions $f$ and $g$

\begin{equation}\label{eq fg}
\smo{f \, g} \eq f_l \, g_l + \smo{f_s g_s} + \frac{1}{\Lambda^2} \vec{\nabla}f_l \cdot \vec{\nabla}g_l + \mbox{higher derivative terms} \;.
\end{equation}
where we assumed the smoothing function to be a Gaussian and the smoothing scale $\Lambda$ has been defined in such a way that $\int d^3x W_\Lambda(\vec{x}) \, x^i x^j = \delta^{ij}/\Lambda^2$. The terms that have been neglected are terms with more than one derivative acting on the smoothed functions $f_l$ and $g_l$. For our discussion of the velocity, we need the analogous formula for the ratio of two functions rather than their product. First, note that for some function $g$ the smoothed inverse can be expanded in terms of derivatives of $g_l$  

\begin{equation}
\smo{\frac{1}{g}} \eq \frac{1}{g_l} - \frac{\vec{\nabla} g_l}{g_l^2} \cdot \smo{ \frac{\vec{x}-\vec{y}}{1+\frac{g_s(\vec{y})}{g_l(\vec{x})}} } + \mathcal{O}(\partial^2 g_l)\;,
\end{equation}
where in a small abuse of notation we denote $\vec{y}$ as the integration variable of the smoothing integral. Second, we can use the fact that any smoothed function has only small fluctuations compared to a constant background value, i.e $g = g_l + g_s = \bar{g} + \delta g + g_s$ where $\delta g/\bar{g}<1 $. This allows us to factorize the pure short-scale contributions since we now have

\begin{equation}
\smo{\frac{f_s}{g} } \eq \smo{ \frac{f_s}{\bar{g}+g_s}} - \delta g \smo{\frac{f_s}{(\bar{g}+g_s)^2} } + \mathcal{O}(\delta g^2) + \mathcal{O}(\partial\, \delta g)
\end{equation}
In the expression above we neglected terms with higher powers of $\delta g$ as well as derivative terms, i.e. $\vec{\nabla}\delta g$. Using the fact that the purely short modes of $1/g$ are given by $1/g - 1/g_l +\mathcal{O}(\partial g_l)$ and that $\smo{f_s/g_l}=0+ \mathcal{O}(\partial g_l)$, we can use Eq.~\eqref{eq fg} we end up with 

\begin{equation}
\smo{\frac{f}{g} } \eq \frac{f_l}{g_l} + \smo{ \frac{f_s}{\bar{g}+g_s}} - \delta g \smo{\frac{f_s}{(\bar{g}+g_s)^2} } + \mathcal{O}(\delta g^2) + \mathcal{O}(\partial \smo{..} ) \;.
\end{equation}
where again we neglected derivatives on smoothed quantities as well as higher powers of $\delta g$ and we did not expand the first term, i.e.~$f_l/g_l$ in powers of $\delta g$. Note well that now the smoothing is only taken over pure short scale quantities.

   
\subsection{Volume-weighted velocity}\label{a:vwv}

In the literature one finds two different ways to define a velocity for a system of $N$ particles (see e.g.~\cite{Bernardeau1995,Bernardeau1996} for a discussion). If one knows the velocity of each of the particles, one can define some velocity field $\vdis$ by smooth interpolation. Then, it is possible to define the \textit{volume-weighted velocity} to be 
\be\label{vwv}
\vvol(\vec x)\equiv \frac{\int d^3 y \,\vdis(\vec y) W_{\rm vol}(\vec x- \vec y)}{\int d^3 y \,  W_{\rm vol}(\vec x- \vec y)}\,,
\ee
where $W_{\rm vol}$ is some filter function with support in some compact region (e.g.~a Gaussian). A different definition is obtained when one uses also the information about the local density. One defines the \textit{mass-weighted velocity} to be 
\be\label{mwv}
\vmas(\vec x)\equiv \frac{\int d^3 y \, \vdis(\vec y) \, \rho(\vec y)\, W_{\rm mass}(\vec x- \vec y)}{\int d^3 y\, \rho(\vec y)\,   W_{\rm mass}(\vec x- \vec y)}\,,
\ee
where again $W_{\rm mass}$ is some filter function. Notice that $\vdis \rho$ is the local momentum. In words, the above expression says that the mass-weighted velocity is obtained by smoothing the momentum field and \textit{then} dividing by the smoothed density field. But this is precisely the definition that is commonly used for the velocity in the EFTofLSS, where we defined $\vec v_{l} \equiv\smo{\bm \pi}/\smo{\rho}$ plus counterterms to cancel the cutoff dependence of the composite operator (see Sec.~\ref{s:sims}). However, one could have also formally\footnote{In practice this is hard to implement because the unsmoothed density can be a very complicated object, including for example caustics.} defined $\tilde{  v_{l}} \equiv \smo{ \bm \pi/\rho }$, again up to counterterms. In the EFT approach, in which the smoothing plays explicitly a crucial role, it is clear what the difference between these two definition is. Using some useful formulae derived in Appx.~\ref{a:smooth} and up to counterterms, one finds 

\begin{equation}
\begin{split}\label{rela}
\tilde{\vec{v}}_l \eq & \vec{v}_l + \frac{1}{\bar{\rho}} \smo{\frac{\bm{\pi}_s}{1 + \delta_s}} - \frac{\delta_l}{\bar{\rho}} \smo{\frac{\bm{\pi}_s}{(1 + \delta_s)^2} } + \mbox{higher order terms} \;, \\[1.5ex]
\end{split}
\end{equation}
where the subscript $\delta_s$ and $\bm{\pi}_s$ represent the pure short distance fluctuations of $\delta = \rho/\bar{\rho}- 1$ and $\bm{\pi}$. The terms that have been neglected are terms with derivatives acting on smoothed quantities as well as higher powers of $\delta_l$. The volume- and mass-weighted velocities hence differ by a pure short-scale term plus higher order terms. We refrain from going into the details of the definition of the volume-weighted velocity field since this involves technical difficulties as discussed e.g.~in Ref.~\cite{Bernardeau1995}. 

One interesting property of the mass-weighted velocity is that it must obey some particular relation as consequence of momentum conservation (see Appx.~\ref{sec JJ}). Also, the mass-weighted velocity is more readily extracted from simulations (or observations) by simply summing over the momenta of the particles in a given region and dividing by its total mass. This is to be contrasted with the volume-weighted velocity that, in order to be extracted from an $N$-body simulation, requires the introduction of some tessellation, e.g.~along the lines of Ref.~\cite{Bernardeau1995}. Note that because of the relation \eqref{rela}, mass- and volume-weighted velocities are not perfectly correlated, but rather differ by at least some noise term. Some hints of this could be seen in Fig.~3 of Ref.~\cite{Bernardeau1996}, although a thorough comparison with simulations requires further investigation and is left for future work.


\subsection{Analytic expressions for $P_{\delta \theta}$ and $P_{\theta\theta}$ in dim reg} \label{sec formulae}

Let us collect the full expressions of the one-loop power spectra of $P_{\delta \theta} $ and $P_{\theta \theta}$. A \emph{Mathematica} file containing the following expressions is attached to the arXiv.org submission of this paper.

The coefficients $\alpha_{1,\delta \theta}$, $\alpha_{1, \theta \theta}$ and $\tilde{\alpha}_{\delta \theta}$, $\alpha_{ \theta \theta}$ can be easily deduced from the full one-loop contributions to the power spectra $P_{\delta \theta}$ and $P_{\theta \theta}$ for an arbitrary spectral index $n$ and space dimension $d$. We use dimensional regularization in order to regularize the loop integrals and adopt the following convention for the two contributions

\begin{equation}\label{P22all}
\begin{split}
P_{22, O_1 O_2} (k,\tau) \eq & 2 (-\hubble)^{n_\theta} \int \frac{d^d q}{(2\pi)^d} \; P_\textit{lin}(\mathbf{q}) P_\textit{lin}(\mathbf{k}-\mathbf{q}) \, K_2^{(O_1)}(\mathbf{k}-\mathbf{q},\mathbf{q}) K_2^{(O_2)}(\mathbf{k}-\mathbf{q},\mathbf{q})  \\[1.5ex]
P_{13, O_1 O_2} (k,\tau) \eq &   3 (-\hubble)^{n_\theta} P_\textit{lin}(\mathbf{k} ) \int \frac{d^d q}{(2\pi)^d} \;   P_\textit{lin}(\mathbf{q})   \left\{ K_3^{(O_1)}(\mathbf{k},\mathbf{q},-\mathbf{q}) + K_3^{(O_2)}(\mathbf{k},\mathbf{q},-\mathbf{q}) \right\} \;,
\end{split}
\end{equation} 
where $O_1$ and $O_2$ stand for either $\delta$ or $\theta$ and $n_\theta$ is the number of $\theta$ among $O_1$ and $O_2$. Note the sum of the two kernels in the definition of the correlators in Eq.~\ref{P22all} (in the main text we often discuss the two contributions separately). The kernel $K_m^{(O_i)}(\mathbf{k}_1,.. \mathbf{k}_m)$ is simply the SPT kernel function $F_m(\mathbf{k}_1,... \mathbf{k}_m)$ or $G_m(\mathbf{k}_1,... \mathbf{k}_m)$ of Eq.~\eqref{eq loop} depending on $O_i$. The linear power spectrum in $d$-dimension is given by $P_\textit{lin}(k,\tau) = a^{2d-4} P_\textit{in} = a^{2d-4} A k^n$ as discussed in the appendix of Ref.~\cite{Pajer2013} where one can also find the expressions for $P_{22,\delta\delta}$ and $P_{13,\delta\delta}$ \footnote{Note that in Ref.~\cite{Pajer2013} the quantities $P_{22,\delta\delta}$ and $P_{13,\delta\delta}$ have been defined with a factor $1/(2\pi)^3$ rather than $1/(2\pi)^d$}. For the correlators involving the velocity divergence, the results read

\begin{equation}
\begin{split}
P_{22,\delta \theta}(k,\tau)  \eq & - \hubble \,  2^{-d-2} \pi ^{-\frac{d}{2}} A^2 a^{4(d-2)} k^{d+2n} \, \Bigg\{ \frac{4 \Gamma \left(-\frac{d}{2}-n+4\right) \Gamma \left(\frac{1}{2} (d+n-4)\right)^2}{49 \Gamma
   \left(2-\frac{n}{2}\right)^2 \Gamma (d+n-4)} \\
& + \frac{10 \Gamma \left(-\frac{d}{2}-n+3\right) \Gamma
   \left(\frac{1}{2} (d+n-2)\right) \Gamma \left(\frac{1}{2} (d+n-4)\right)}{49 \Gamma
   \left(1-\frac{n}{2}\right) \Gamma \left(2-\frac{n}{2}\right) \Gamma (d+n-3)} \\
& - \frac{29 \Gamma \left(-\frac{d}{2}-n+2\right) \Gamma \left(\frac{d+n}{2}\right) \Gamma \left(\frac{1}{2}
   (d+n-4)\right)}{49 \Gamma \left(2-\frac{n}{2}\right) \Gamma \left(-\frac{n}{2}\right) \Gamma
   (d+n-2)} \\
& -\frac{4 \Gamma \left(-\frac{d}{2}-n+1\right) \Gamma \left(\frac{1}{2} (d+n+2)\right) \Gamma
   \left(\frac{1}{2} (d+n-4)\right)}{49 \Gamma \left(-\frac{n}{2}-1\right) \Gamma \left(2-\frac{n}{2}\right)
   \Gamma (d+n-1)} \\
& +\frac{15 \Gamma \left(-\frac{d}{2}-n\right) \Gamma \left(\frac{1}{2} (d+n+4)\right) \Gamma
   \left(\frac{1}{2} (d+n-4)\right)}{49 \Gamma \left(-\frac{n}{2}-2\right) \Gamma \left(2-\frac{n}{2}\right)
   \Gamma (d+n)} \\
& +\frac{4 \Gamma \left(-\frac{d}{2}-n+1\right) \Gamma \left(\frac{1}{2} (d+n-2)\right) \Gamma
   \left(\frac{d+n}{2}\right)}{49 \Gamma \left(1-\frac{n}{2}\right) \Gamma \left(-\frac{n}{2}\right) \Gamma
   (d+n-1)} \\
& -\frac{60 \Gamma \left(-\frac{d}{2}-n\right) \Gamma \left(\frac{1}{2} (d+n-2)\right) \Gamma
   \left(\frac{1}{2} (d+n+2)\right)}{49 \Gamma \left(-\frac{n}{2}-1\right) \Gamma \left(1-\frac{n}{2}\right)
   \Gamma (d+n)} \\
& +\frac{23 \Gamma \left(-\frac{d}{2}-n+2\right) \Gamma \left(\frac{1}{2} (d+n-2)\right)^2}{49
   \Gamma \left(1-\frac{n}{2}\right)^2 \Gamma (d+n-2)}+\frac{45 \Gamma \left(-\frac{d}{2}-n\right) \Gamma
   \left(\frac{d+n}{2}\right)^2}{49 \Gamma \left(-\frac{n}{2}\right)^2 \Gamma (d+n)}  \Bigg\}  \;, 
\end{split}
\end{equation}

\begin{equation}
\begin{split}
P_{13,\delta \theta}(k,\tau) \eq & - \hubble \,  2^{-d-2} \pi ^{-\frac{d}{2}} A^2 a^{4(d-2)} k^{d+2n} \,\Bigg\{  -\frac{4 \Gamma \left(\frac{d}{2}-1\right) \Gamma \left(-\frac{d}{2}-\frac{n}{2}+3\right) \Gamma
   \left(\frac{1}{2} (d+n-4)\right)}{21 \Gamma \left(2-\frac{n}{2}\right) \Gamma
   \left(d+\frac{n}{2}-3\right)} \\
& +\frac{\Gamma \left(\frac{d}{2}-1\right) \Gamma
   \left(-\frac{d}{2}-\frac{n}{2}+2\right) \Gamma \left(\frac{1}{2} (d+n-2)\right)}{3 \Gamma
   \left(1-\frac{n}{2}\right) \Gamma \left(d+\frac{n}{2}-2\right)} \\
& +\frac{\Gamma \left(\frac{d}{2}-1\right)
   \Gamma \left(-\frac{d}{2}-\frac{n}{2}+1\right) \Gamma \left(\frac{d+n}{2}\right)}{7 \Gamma
   \left(-\frac{n}{2}\right) \Gamma \left(d+\frac{n}{2}-1\right)} \\
& -\frac{11 \Gamma \left(\frac{d}{2}-1\right)
   \Gamma \left(-\frac{d}{2}-\frac{n}{2}\right) \Gamma \left(\frac{1}{2} (d+n+2)\right)}{21 \Gamma
   \left(-\frac{n}{2}-1\right) \Gamma \left(d+\frac{n}{2}\right)} \\
& +\frac{5 \Gamma \left(\frac{d}{2}-1\right)
   \Gamma \left(-\frac{d}{2} -\frac{n}{2}-1\right) \Gamma \left(\frac{1}{2} (d+n+4)\right)}{21 \Gamma
   \left(-\frac{n}{2}-2\right) \Gamma \left(d+\frac{n}{2}+1\right)}   \Bigg\} \;,
\end{split}
\end{equation}

\begin{equation}
\begin{split}
P_{22,\theta \theta}(k,\tau) \eq &   \hubble^2 \,  2^{-d-2} \pi ^{-\frac{d}{2}} A^2 a^{4(d-2)} k^{d+2n} \, \Bigg\{ \frac{8 \Gamma \left(-\frac{d}{2}-n+4\right) \Gamma \left(\frac{1}{2} (d+n-4)\right)^2}{49 \Gamma
   \left(2-\frac{n}{2}\right)^2 \Gamma (d+n-4)} \\
& -\frac{8 \Gamma \left(-\frac{d}{2}-n+3\right) \Gamma
   \left(\frac{1}{2} (d+n-2)\right) \Gamma \left(\frac{1}{2} (d+n-4)\right)}{49 \Gamma
   \left(1-\frac{n}{2}\right) \Gamma \left(2-\frac{n}{2}\right) \Gamma (d+n-3)} \\
& -\frac{23 \Gamma \left(-\frac{d}{2}-n+2\right) \Gamma \left(\frac{d+n}{2}\right) \Gamma \left(\frac{1}{2}
   (d+n-4)\right)}{49 \Gamma \left(2-\frac{n}{2}\right) \Gamma \left(-\frac{n}{2}\right) \Gamma
   (d+n-2)} \\
& +\frac{6 \Gamma \left(-\frac{d}{2}-n+1\right) \Gamma \left(\frac{1}{2} (d+n+2)\right) \Gamma
   \left(\frac{1}{2} (d+n-4)\right)}{49 \Gamma \left(-\frac{n}{2}-1\right) \Gamma \left(2-\frac{n}{2}\right)
   \Gamma (d+n-1)} \\
& +\frac{9 \Gamma \left(-\frac{d}{2}-n\right) \Gamma \left(\frac{1}{2} (d+n+4)\right) \Gamma
   \left(\frac{1}{2} (d+n-4)\right)}{49 \Gamma \left(-\frac{n}{2}-2\right) \Gamma \left(2-\frac{n}{2}\right)
   \Gamma (d+n)} \\
& -\frac{6 \Gamma \left(-\frac{d}{2}-n+1\right) \Gamma \left(\frac{1}{2} (d+n-2)\right) \Gamma
   \left(\frac{d+n}{2}\right)}{49 \Gamma \left(1-\frac{n}{2}\right) \Gamma \left(-\frac{n}{2}\right) \Gamma
   (d+n-1)} \\
&-\frac{36 \Gamma \left(-\frac{d}{2}-n\right) \Gamma \left(\frac{1}{2} (d+n-2)\right) \Gamma
   \left(\frac{1}{2} (d+n+2)\right)}{49 \Gamma \left(-\frac{n}{2}-1\right) \Gamma \left(1-\frac{n}{2}\right)
   \Gamma (d+n)} \\
& +\frac{25 \Gamma \left(-\frac{d}{2}-n+2\right) \Gamma \left(\frac{1}{2} (d+n-2)\right)^2}{49
   \Gamma \left(1-\frac{n}{2}\right)^2 \Gamma (d+n-2)} +\frac{27 \Gamma \left(-\frac{d}{2}-n\right) \Gamma
   \left(\frac{d+n}{2}\right)^2}{49 \Gamma \left(-\frac{n}{2}\right)^2 \Gamma (d+n)}  \Bigg\} \;,
\end{split}
\end{equation}

\begin{equation}
\begin{split}
P_{13,\theta \theta}(k,\tau) \eq & \hubble^2 \,  2^{-d-2} \pi ^{-\frac{d}{2}} A^2 a^{4(d-2)} k^{d+2n} \, \Bigg\{  -\frac{2 \Gamma \left(\frac{d}{2}-1\right) \Gamma \left(-\frac{d}{2}-\frac{n}{2}+3\right) \Gamma
   \left(\frac{1}{2} (d+n-4)\right)}{7 \Gamma \left(2-\frac{n}{2}\right) \Gamma
   \left(d+\frac{n}{2}-3\right)} \\
& +\frac{5 \Gamma \left(\frac{d}{2}-1\right) \Gamma
   \left(-\frac{d}{2}-\frac{n}{2}+2\right) \Gamma \left(\frac{1}{2} (d+n-2)\right)}{7 \Gamma
   \left(1-\frac{n}{2}\right) \Gamma \left(d+\frac{n}{2}-2\right)} \\
& -\frac{3 \Gamma \left(\frac{d}{2}-1\right)
   \Gamma \left(-\frac{d}{2}-\frac{n}{2}+1\right) \Gamma \left(\frac{d+n}{2}\right)}{7 \Gamma
   \left(-\frac{n}{2}\right) \Gamma \left(d+\frac{n}{2}-1\right)} \\
& -\frac{\Gamma \left(\frac{d}{2}-1\right)
   \Gamma \left(-\frac{d}{2}-\frac{n}{2}\right) \Gamma \left(\frac{1}{2} (d+n+2)\right)}{7 \Gamma
   \left(-\frac{n}{2}-1\right) \Gamma \left(d+\frac{n}{2}\right)} \\
& +\frac{\Gamma \left(\frac{d}{2}-1\right)
   \Gamma \left(-\frac{d}{2}-\frac{n}{2}-1\right) \Gamma \left(\frac{1}{2} (d+n+4)\right)}{7 \Gamma
   \left(-\frac{n}{2}-2\right) \Gamma \left(d+\frac{n}{2}+1\right)}   \Bigg\} \;.
\end{split}
\end{equation}

 
\section{Subtleties in dimensional regularization}\label{a:dimreg}

In order to extract the coefficients in table \ref{tab alpha} from the formulae given in the previous appendix, one has to expand the Gamma functions around $d=3+2\epsilon$ (the factor of 2 is just a convention) for some fixed $n$. Some subtleties arise when the Gamma functions are divergent in the three-dimensional limit. In these cases, the poles in the Gamma functions are cancelled by expanding the $k^{d}$ (obtained from performing the integral) to linear order in $\epsilon$, hence leading to the logarithmic term in the final correlator. One might wonder whether one should compute the three-dimensional limit of $P_{22}$ or of $P_{22}k^{d}$, the latter being the self-similarity invariant quantity in $d$ dimensions. The final value of the logarithmic coefficients $\tilde \alpha_1$ in Tab.~\ref{tab alpha} would change by a factor of 2. The right thing to do is to use just $P_{22}$ as done in \cite{Pajer2013}. There are three ways to see that this gives the correct result. One way is to simply compute the logarithmic terms using a cutoff regularization. This is pretty straightforward and agrees with the finite part of $P_{22}$ from dimensional regularization (as opposed to $P_{22} k^{d}$). A second way is to compare with simulations. We have done that using data for $n=-1$ kindly provided to us by C.~Orban. By comparing the best fit to the density power spectrum (fitting $\beta$ in \eqref{eq generalI} and neglecting $\gamma$ and higher derivatives) using the $\tilde \alpha_1$ in Tab.~\ref{tab alpha} rather than twice those numbers, one finds that the fit significantly improves (in the statistical sense), especially at large scales where higher order corrections should be negligible. Lastly, from the most practical point of view, dimensional regularization is just a mathematical trick to extract the logarithmically divergent part of some divergent integral\footnote{Of course dimensional regularization extracts also the rest of the divergence and the finite part, but the divergence is cancelled by the counter-term to define the renormalized theory and the finite part is degenerate with the choice of renormalization scheme.}. Clearly the logarithmic divergence is a property of the integral itself and should not depend on what the integral is multiplied by. In other words, the factors of $k^{d}$ in dimensional regularization stem only from the measure of the $d$-dimensional integral, of which at the one-loop level there is only one.


\bibliographystyle{utphys}

\bibliography{EFToLSS_Bibliography.bib}

\providecommand{\href}[2]{#2}\begingroup\raggedright\begin{thebibliography}{10}

\bibitem{Weinberg1996a}
S.~Weinberg,
``{The quantum theory of fields. Vol. 2: Modern applications},''.

\bibitem{Bernardeau2002}
F.~Bernardeau, S.~Colombi, E.~Gaztanaga, and R.~Scoccimarro, ``{Large scale
  structure of the universe and cosmological perturbation theory},''
  \href{http://dx.doi.org/10.1016/S0370-1573(02)00135-7}{{\em Phys.Rept.}
  {\bfseries 367} (2002) 1--248},
\href{http://arxiv.org/abs/astro-ph/0112551}{{\ttfamily arXiv:astro-ph/0112551
  [astro-ph]}}.

\bibitem{Baumann2012}
D.~Baumann, A.~Nicolis, L.~Senatore, and M.~Zaldarriaga, ``{Cosmological
  Non-Linearities as an Effective Fluid},''
  \href{http://dx.doi.org/10.1088/1475-7516/2012/07/051}{{\em JCAP} {\bfseries
  1207} (2012) 051},
\href{http://arxiv.org/abs/1004.2488}{{\ttfamily arXiv:1004.2488
  [astro-ph.CO]}}.

\bibitem{Carrasco2012}
J.~J.~M. Carrasco, M.~P. Hertzberg, and L.~Senatore, ``{The Effective Field
  Theory of Cosmological Large Scale Structures},''
  \href{http://dx.doi.org/10.1007/JHEP09(2012)082}{{\em JHEP} {\bfseries 1209}
  (2012) 082},
\href{http://arxiv.org/abs/1206.2926}{{\ttfamily arXiv:1206.2926
  [astro-ph.CO]}}.

\bibitem{Hertzberg2012}
M.~P. Hertzberg, ``{The Effective Field Theory of Dark Matter and Structure
  Formation: Semi-Analytical Results},''
\href{http://arxiv.org/abs/1208.0839}{{\ttfamily arXiv:1208.0839
  [astro-ph.CO]}}.

\bibitem{Pajer2013}
E.~Pajer and M.~Zaldarriaga, ``{On the Renormalization of the Effective Field
  Theory of Large Scale Structures},''
\href{http://arxiv.org/abs/1301.7182}{{\ttfamily arXiv:1301.7182
  [astro-ph.CO]}}.

\bibitem{Carrasco2013}
J.~J.~M. Carrasco, S.~Foreman, D.~Green, and L.~Senatore, ``{The 2-loop matter
  power spectrum and the IR-safe integrand},''
\href{http://arxiv.org/abs/1304.4946}{{\ttfamily arXiv:1304.4946
  [astro-ph.CO]}}.

\bibitem{Peebles1980}
{J.~P.~E.~Peebles}, {\em {The large scale structure of the universe}}.
\newblock {Princeton University Press}, 1980.

\bibitem{Zeldovich1965}
{Ya.~B.~Zel'dovich} {\em Adv. Astron. Astrophys.} {\bfseries 3} (1965) 241.

\bibitem{Goroff1986}
M.~Goroff, B.~Grinstein, S.~Rey, and M.~B. Wise, ``{Coupling of Modes of
  Cosmological Mass Density Fluctuations},''
\href{http://dx.doi.org/10.1086/164749}{{\em Astrophys.J.} {\bfseries 311}
  (1986) 6--14}.

\bibitem{Crocce2008}
M.~Crocce and R.~Scoccimarro, ``{Nonlinear Evolution of Baryon Acoustic
  Oscillations},'' \href{http://dx.doi.org/10.1103/PhysRevD.77.023533}{{\em
  Phys.Rev.} {\bfseries D77} (2008) 023533},
\href{http://arxiv.org/abs/0704.2783}{{\ttfamily arXiv:0704.2783 [astro-ph]}}.

\bibitem{Pueblas2009}
S.~Pueblas and R.~Scoccimarro, ``{Generation of Vorticity and Velocity
  Dispersion by Orbit Crossing},''
  \href{http://dx.doi.org/10.1103/PhysRevD.80.043504}{{\em Phys.Rev.}
  {\bfseries D80} (2009) 043504},
\href{http://arxiv.org/abs/0809.4606}{{\ttfamily arXiv:0809.4606 [astro-ph]}}.

\bibitem{Valageas2011}
P.~Valageas, ``{Impact of shell crossing and scope of perturbative approaches
  in real and redshift space},'' {\em Astron.Astrophys.} {\bfseries 526} (2011)
  A67,
\href{http://arxiv.org/abs/1009.0106}{{\ttfamily arXiv:1009.0106
  [astro-ph.CO]}}.

\bibitem{Pietroni2012}
M.~Pietroni, G.~Mangano, N.~Saviano, and M.~Viel, ``{Coarse-Grained
  Cosmological Perturbation Theory},''
  \href{http://dx.doi.org/10.1088/1475-7516/2012/01/019}{{\em JCAP} {\bfseries
  1201} (2012) 019},
\href{http://arxiv.org/abs/1108.5203}{{\ttfamily arXiv:1108.5203
  [astro-ph.CO]}}.

\bibitem{Crocce2006a}
M.~Crocce and R.~Scoccimarro, ``{Memory of initial conditions in gravitational
  clustering},'' \href{http://dx.doi.org/10.1103/PhysRevD.73.063520}{{\em
  Phys.Rev.} {\bfseries D73} (2006) 063520},
\href{http://arxiv.org/abs/astro-ph/0509419}{{\ttfamily arXiv:astro-ph/0509419
  [astro-ph]}}.

\bibitem{Crocce2006}
M.~Crocce and R.~Scoccimarro, ``{Renormalized cosmological perturbation
  theory},'' \href{http://dx.doi.org/10.1103/PhysRevD.73.063519}{{\em
  Phys.Rev.} {\bfseries D73} (2006) 063519},
\href{http://arxiv.org/abs/astro-ph/0509418}{{\ttfamily arXiv:astro-ph/0509418
  [astro-ph]}}.

\bibitem{Matarrese2007}
S.~Matarrese and M.~Pietroni, ``{Resumming Cosmic Perturbations},''
  \href{http://dx.doi.org/10.1088/1475-7516/2007/06/026}{{\em JCAP} {\bfseries
  0706} (2007) 026},
\href{http://arxiv.org/abs/astro-ph/0703563}{{\ttfamily arXiv:astro-ph/0703563
  [astro-ph]}}.

\bibitem{Sugiyama2013}
N.~S. Sugiyama and D.~N. Spergel, ``{How does non-linear dynamics affect the
  baryon acoustic oscillation?},''
\href{http://arxiv.org/abs/1306.6660}{{\ttfamily arXiv:1306.6660
  [astro-ph.CO]}}.

\bibitem{Weinberg:2008zzc}
S.~Weinberg,
``{Cosmology},''.

\bibitem{Bernardeau1995}
F.~Bernardeau and R.~van~de Weygaert, ``{A New method for accurate velocity
  statistics estimation},''
\href{http://arxiv.org/abs/astro-ph/9508125}{{\ttfamily arXiv:astro-ph/9508125
  [astro-ph]}}.

\bibitem{Bernardeau1996}
F.~Bernardeau, ``{The Statistics of the large scale velocity field},''
\href{http://arxiv.org/abs/astro-ph/9601083}{{\ttfamily arXiv:astro-ph/9601083
  [astro-ph]}}.

\bibitem{Carrasco2013a}
J.~J.~M. Carrasco, S.~Foreman, D.~Green, and L.~Senatore, ``{The Effective
  Field Theory of Large Scale Structures at Two Loops},''
\href{http://arxiv.org/abs/1310.0464}{{\ttfamily arXiv:1310.0464
  [astro-ph.CO]}}.

\bibitem{Scoccimarro2000}
R.~Scoccimarro, ``{A new angle on gravitational clustering},''
\href{http://arxiv.org/abs/astro-ph/0008277}{{\ttfamily arXiv:astro-ph/0008277
  [astro-ph]}}.

\bibitem{Jain1996}
B.~Jain and E.~Bertschinger, ``{Selfsimilar evolution of cosmological density
  fluctuations},'' \href{http://dx.doi.org/10.1086/176625}{{\em Astrophys.J.}
  {\bfseries 456} (1996) 43},
\href{http://arxiv.org/abs/astro-ph/9503025}{{\ttfamily arXiv:astro-ph/9503025
  [astro-ph]}}.

\bibitem{Scoccimarro1996}
R.~Scoccimarro and J.~Frieman, ``{Loop corrections in nonlinear cosmological
  perturbation theory},'' \href{http://dx.doi.org/10.1086/192306}{{\em
  Astrophys.J.Suppl.} {\bfseries 105} (1996) 37},
\href{http://arxiv.org/abs/astro-ph/9509047}{{\ttfamily arXiv:astro-ph/9509047
  [astro-ph]}}.

\bibitem{Kehagias2013}
A.~Kehagias and A.~Riotto, ``{Symmetries and Consistency Relations in the Large
  Scale Structure of the Universe},''
  \href{http://dx.doi.org/10.1016/j.nuclphysb.2013.05.009}{{\em Nucl.Phys.}
  {\bfseries B873} (2013) 514--529},
\href{http://arxiv.org/abs/1302.0130}{{\ttfamily arXiv:1302.0130
  [astro-ph.CO]}}.

\bibitem{Peloso2013}
M.~Peloso and M.~Pietroni, ``{Galilean invariance and the consistency relation
  for the nonlinear squeezed bispectrum of large scale structure},''
  \href{http://dx.doi.org/10.1088/1475-7516/2013/05/031}{{\em JCAP} {\bfseries
  1305} (2013) 031},
\href{http://arxiv.org/abs/1302.0223}{{\ttfamily arXiv:1302.0223
  [astro-ph.CO]}}.

\bibitem{Blas2013}
D.~Blas, M.~Garny, and T.~Konstandin, ``{On the non-linear scale of
  cosmological perturbation theory},''
\href{http://arxiv.org/abs/1304.1546}{{\ttfamily arXiv:1304.1546
  [astro-ph.CO]}}.

\end{thebibliography}\endgroup

\addcontentsline{toc}{section}{References}

\end{document}